\documentclass{ieeeaccess}

\usepackage{amsmath}
\usepackage{amssymb}
\usepackage{bm}
\usepackage{cite}
\usepackage{graphicx}
\graphicspath{{images/}}
\usepackage[caption=false,font=footnotesize,labelfont=rm,textfont=rm]{subfig}
\usepackage{booktabs}
\usepackage{tabularx}
\usepackage[final]{microtype}
\usepackage{textcomp}
\usepackage{makecell}
\usepackage{multirow}
\usepackage{siunitx}
\usepackage{xfp}
\usepackage{flafter}
\usepackage{stfloats}
\usepackage[section]{placeins}
\usepackage{float}
\usepackage{balance}
\DeclareSIUnit{\pu}{\text{p.u.}}
\allowdisplaybreaks[1]

\newcommand{\code}[1]{\texttt{#1}}
\newcommand{\rTwoDisplay}[1]{\num[minimum-decimal-digits=2,round-mode=places,round-precision=2]{\fpeval{abs((#1)-1)<1e-12 ? 1 : ((#1<1 && round((#1),2)>=1) ? 0.99 : (#1))}}}

\begin{document}

\history{}
\doi{}

\title{PowerModelsGAT-AI: Physics-Informed Graph Attention for Multi-System Power Flow with Continual Learning}

\author{\uppercase{Chidozie~Ezeakunne}\authorrefmark{1}$^{,}$\authorrefmark{2},
\uppercase{Jose~E.~Tabarez}\authorrefmark{1},
\uppercase{Reeju~Pokharel}\authorrefmark{1}, and
\uppercase{Anup~Pandey}\authorrefmark{1}}

\address[1]{Los Alamos National Laboratory, Los Alamos, NM, USA}
\address[2]{Department of Physics, University of Central Florida, Orlando, FL 32816 USA}
\tfootnote{Research presented in this work was supported by the Laboratory Directed Research and Development program of Los Alamos National Laboratory under project number 20250854ECR.}
\markboth
{Ezeakunne \headeretal: PowerModelsGAT-AI for Multi-System Power Flow}
{Ezeakunne \headeretal: PowerModelsGAT-AI for Multi-System Power Flow}
\corresp{Corresponding author: Chidozie Ezeakunne (e-mail: cezeakunne@lanl.gov).}

\begin{abstract}
    Solving the alternating current power flow equations in real time is essential for secure grid operation, yet classical Newton--Raphson solvers can be slow under stressed conditions. Existing graph neural networks for power flow are typically trained on a single system and often degrade on different systems. We present PowerModelsGAT-AI, a physics-informed graph attention network that predicts bus voltages and generator injections. The model uses bus-type-aware masking to handle different bus types and balances multiple loss terms, including a power-mismatch penalty, using learned weights. We evaluate the model on 14 benchmark systems (4 to 6,470 buses) and train a unified model on 13 of these under $N\!-\!2$ (two-branch outage) conditions, achieving an average normalized mean absolute error of $0.89\%$ for voltage magnitudes and $R^2 > 0.99$ for voltage angles. We also show continual learning: when adapting a base model to a new 1,354-bus system, standard fine-tuning causes severe forgetting with error increases exceeding $1000\%$ on base systems, while our experience replay and elastic weight consolidation strategy keeps error increases below $2\%$ and in some cases improves base-system performance. Interpretability analysis shows that learned attention weights correlate with physical branch parameters (susceptance: $r = 0.38$; thermal limits: $r = 0.22$), and feature importance analysis supports that the model captures established power flow relationships.
\end{abstract}


\maketitle

\section{Introduction}

\IEEEPARstart{T}{he} alternating current (AC) power flow problem is a foundational computational tool for secure and economic power system operation and planning. The AC power flow equations describe the nonlinear relationship between bus voltages, power injections, and network admittances, and are fundamental to standalone power flow analysis, optimal power flow (OPF), and security-constrained formulations~\cite{CARPENTIER19793,frank2016introduction,bose2011optimal}. The goal is to determine unknown bus voltages and power injections from specified inputs, which vary by bus type (Table~\ref{tab:bus-masks}).

\begin{table}[t]
    \centering
    \caption{Bus-Type Unknown Targets and Supervision Masks.}
    \label{tab:bus-masks}
    \renewcommand{\arraystretch}{1.15}
    \footnotesize
    \begin{tabular*}{\columnwidth}{@{\extracolsep{\fill}} l c c c @{}}
        \toprule
        \textbf{Bus Type} & \makecell[c]{\textbf{Known in} \\ $\mathbf{y}_i$} & \makecell[c]{\textbf{Unknown} \\ \textbf{Targets}} & \makecell[c]{\textbf{Mask} \\ $[V_m,\delta,P_g,Q_g]$} \\
        \midrule
        PQ (Load)         & $P_g, Q_g$                & $V_m, \delta$        & [1, 1, 0, 0]                           \\
        PV (Generator)    & $P_g, V_m$                & $\delta, Q_g$        & [0, 1, 0, 1]                           \\
        Slack             & $V_m, \delta$             & $P_g, Q_g$           & [0, 0, 1, 1]                           \\
        \bottomrule
    \end{tabular*}
\end{table}

For a system with $N$ buses, let $V_i = V_{m,i} e^{j\delta_i}$ be the complex voltage at bus $i$, where $V_{m,i}$ is the voltage magnitude, $\delta_i$ is the voltage angle, and $j=\sqrt{-1}$. Given the bus-admittance matrix $\mathbf{Y} \in \mathbb{C}^{N\times N}$, the net complex power injection $S_i^{\mathrm{inj}}$ at each non-slack bus $i$ satisfies
\begin{equation}
    S_i^{\mathrm{inj}}
    = P_{g,i} - P_{d,i} + j(Q_{g,i} - Q_{d,i})
    = V_i \Big(\sum_{k=1}^{N} Y_{ik} V_k \Big)^{\!*},
    \label{eq:acpf-complex}
\end{equation}
where $P_{g,i}$ and $Q_{g,i}$ are the active and reactive power generation at bus $i$, $P_{d,i}$ and $Q_{d,i}$ are the active and reactive power demand, and $(\cdot)^*$ is the complex conjugate. Decomposing the admittance $Y_{ik} = G_{ik}+jB_{ik}$ into its conductance $G_{ik}$ and susceptance $B_{ik}$, \eqref{eq:acpf-complex} yields the familiar pair of nonlinear equations per non-slack bus:

\begin{align}
    P_{g,i} - P_{d,i} & = V_{m,i} \sum_{k=1}^{N} V_{m,k}\!\left[G_{ik}\cos(\delta_i - \delta_k)\right. \nonumber \\
                      & \qquad \left. + B_{ik}\sin(\delta_i - \delta_k)\right], \label{eq:p_line}                \\
    Q_{g,i} - Q_{d,i} & = V_{m,i} \sum_{k=1}^{N} V_{m,k}\!\left[G_{ik}\sin(\delta_i - \delta_k)\right. \nonumber \\
                      & \qquad \left. - B_{ik}\cos(\delta_i - \delta_k)\right]. \label{eq:q_line}
\end{align}

In standard AC power flow analysis, one solves for the unknown state variables given specified power injections and voltage setpoints. OPF augments these equations with an economic objective and additional constraints~\cite{wood2013power,jiang2024advancements}, but OPF is not the focus of this work.

We explicitly treat the complete per-bus state
\begin{equation}
    \mathbf{y}_i = [V_{m,i},\,\delta_i,\,P_{g,i},\,Q_{g,i}],
\end{equation}
and design PMGAT-AI to infer the \emph{unknown} components of $\mathbf{y}_i$ for each bus type (PQ, PV, and Slack) as summarized in Table~\ref{tab:bus-masks}. Thus, a single model predicts bus voltages ($V_m,\delta$ at PQ buses) and generator injections ($P_g,Q_g$ at slack buses) within a unified AC power flow formulation.

Historically, operators executed AC power flow primarily for offline planning, day-ahead scheduling, and periodic security assessment, for which the computational cost of iterative Newton-type solvers was acceptable~\cite{frank2016introduction,stott1974fast}. This assumption is increasingly strained.

High penetrations of inverter-based renewable energy sources and distributed energy resources introduce faster fluctuations, bidirectional flows, and operating points closer to security boundaries~\cite{mohagheghi2018survey,wang2023ac,feng2024safe}. As a result, AC power flow analysis is becoming a near real-time operational requirement, with operators requiring frequent solutions for contingency analysis and corrective actions~\cite{mohagheghi2018survey,feng2024safe}.

Under these stressed or ill-conditioned conditions, convergence and robustness of Newton--Raphson solvers can degrade~\cite{iwamoto1981load,dehghanpour2018survey}.

Due to the trade-off between computational speed and solution reliability, there has been significant research interest in data-driven, machine-learning-based alternatives for predicting AC power flow, with some extensions to OPF and state estimation~\cite{donon2020neural,lopez2023power,lin2024powerflownet,ugwumadu2025powermodel,hu2024adaptive,taghizadeh2024multi,yang2024probabilistic,bottcher2023solving,deihim2024initial,owerko2024unsupervised,lin2022elegnn}. Earlier ML architectures modeled the problem using engineered features and either multilayer perceptrons or convolutional neural networks, with the downside that they used little or no explicit graph-structure information from power systems~\cite{jiang2024advancements,pineda2025beyond}. This limitation often constrained those models to specific topologies, with limited generalization beyond training conditions and across systems~\cite{pineda2025beyond,khaloie2025review}.

Graph neural networks (GNNs) have since emerged as an effective framework for power grids, with buses as nodes, and physical branches (lines and transformers) represented as edges~\cite{wu2020comprehensive,liao2021review}. Message passing in a GNN mirrors the physical dependence of each bus on its electrical neighbors, and existing works have applied GNNs to AC power flow prediction~\cite{lin2024powerflownet,pham2022neural,talebi2025graph,ugwumadu2025powermodel}, OPF~\cite{owerko2024unsupervised,arowolo2025towards}, contingency analysis~\cite{nakiganda2025graphneuralnetworksfast}, and power system state estimation~\cite{lin2022elegnn}. Meanwhile, physics-informed GNNs for power systems incorporate AC power flow residuals or related constraints as differentiable penalties in the loss function, encouraging physically consistent predictions~\cite{karniadakis2021physics,raissi2019physics,lin2022elegnn,eeckhout2024improved,li2025physics}.

However, four gaps remain:

\begin{enumerate}

    \item \textbf{System-specific models and limited transferability.}
          Most existing GNNs for power flow are trained on a single, fixed system with varied operating conditions, requiring a separate specialized model for each system and contingency regime~\cite{donon2020neural,lopez2023power,lin2024powerflownet}. These models often fail to generalize under topology changes and are impractical for deployment where operators must maintain multiple models across systems and conditions~\cite{pineda2025beyond}.

    \item \textbf{Computational bottleneck in hybrid GNN-solver frameworks.}
          Another line of work combines GNNs with iterative solvers, where the GNN provides a warm start or predictions validated by a physics-based criterion~\cite{deihim2024initial,shamseldein2026hybrid,owerko2024unsupervised,ugwumadu2025powermodel}. However, the speed advantage depends on the robustness of the trained model; frequent solver fallback on stressed or contingency cases diminishes the gain. A highly robust standalone model is therefore a prerequisite for effective hybrid deployment, and this work focuses on improving standalone predictive performance.

    \item \textbf{Catastrophic forgetting in continual learning.}
          Real-world grids are nonstationary, and several works advocate on-the-fly fine-tuning where difficult cases are solved with a trusted AC solver and used to update the model~\cite{ugwumadu2025powermodel,jiang2024advancements}. However, naive fine-tuning can lead to \emph{catastrophic forgetting}, where performance on previously learned tasks collapses~\cite{kirkpatrick2017overcoming,carta2022catastrophic}. This challenge is recognized as open for deep graph networks~\cite{carta2022catastrophic}, yet explicit forgetting mitigation remains limited in published AC power flow GNN studies~\cite{ugwumadu2025powermodel}.

    \item \textbf{Limited physical interpretability.}
          Although interpretability has received some attention in GNN-based power system modeling~\cite{lin2024powerflownet,varbella2024powergraph}, systematic validation that learned representations reflect known electrical principles remains limited. Established explainability methods such as GNNExplainer~\cite{ying2019gnnexplainer} and integrated gradients~\cite{sundararajan2017axiomatic} are available for graph learning, yet without such validation it remains unclear whether predictions rely on physically meaningful features or spurious correlations.

\end{enumerate}

To address these gaps, we introduce PowerModelsGAT-AI (hereafter PMGAT-AI), a physics-informed model for solving AC power flow equations.

\subsection{Contributions}

Our contributions are:

\begin{enumerate}

    \item \textbf{Unified multi-system learning for AC power flow.} We evaluate PMGAT-AI on 14 standard benchmarks (4 to 6,470 buses), including system-specific baselines across all systems and a unified model trained across 13 systems. This establishes a single framework for unified learning across heterogeneous grid topologies and operating conditions.

    \item \textbf{Bus-type-aware prediction formulation.} We formulate AC power flow prediction at the bus level as learning the unknown subset of $\mathbf{y}_i=[V_m,\,\delta,\,P_g,\,Q_g]$ under a bus-type-aware supervision mask (Table~\ref{tab:bus-masks}). This enables one model to jointly infer bus voltages and generator injections without leaking specified inputs into supervised losses.

    \item \textbf{Physics-informed graph-attention training objective.} PMGAT-AI combines edge-aware graph attention with a differentiable power-mismatch constraint derived from \eqref{eq:acpf-complex}. We optimize the four supervised targets and the physics term jointly using graph-normalized losses with homoscedastic uncertainty weighting, avoiding manual loss-weight tuning.

    \item \textbf{Continual learning with forgetting mitigation and physics-informed alignment.} For adaptation to new systems or operating conditions, we integrate experience replay with elastic weight consolidation (EWC) to mitigate catastrophic forgetting. Separately, we analyze learned attention and feature attributions, showing alignment with known electrical relationships.
\end{enumerate}

We organize the rest of the paper as follows. Sec.~\ref{sec:dataset} describes dataset generation and graph construction. Sec.~\ref{sec:gnn-model} presents the model architecture, loss design, and optimization. Sec.~\ref{sec:continual} describes the continual learning framework. Secs.~\ref{sec:setup} and~\ref{sec:results} report the evaluation setup and results. Sec.~\ref{sec:interpretability} presents the interpretability analysis, and we conclude in Sec.~\ref{sec:conclusion}. The appendices provide feature definitions, contingency statistics, detailed per-system results, and continual learning analysis.

\section{Dataset Generation and Preprocessing}\label{sec:dataset}
\noindent We generated a dataset from 14 standard power systems\footnote{All system names follow the \code{pandapower} naming conventions~\cite{pandapower2018, matpower2011}; variations in size and naming (e.g., \code{case4gs}, \code{case14}, \code{case\_illinois200}) reflect the original system identifiers.} spanning different scales using \code{pandapower} for AC power flow simulations~\cite{pandapower2018}. These systems enable cross-system training and evaluation. For each system, we extract bus-state vectors $\mathbf{y}_i=[V_m,\delta,P_g,Q_g]$, per-bus input features $\mathbf{x}_i$, and branch-level edge features $\mathbf{e}_{ij}$ (feature definitions in Appendix~\ref{app:features}); bus-type masks define which components are supervised. A full list of the systems considered and contingency statistics are provided in Appendix~\ref{app:contingency-stats}~\cite{matpower2011,josz2016acpf,fliscounakis2013contingency}.

\subsection{Systems and Contingencies}
\noindent We generate scenarios by sampling diverse operating conditions and topologies. To prevent overfitting to global load correlations, we use a hierarchical randomization scheme for active power loads. The load at bus $i$ is scaled by $\rho_i = \gamma \cdot R_{r(i)} \cdot J_i$, where:
\begin{itemize}
    \item $\gamma \sim U[0.7, 1.3]$ is a global scaling factor.
    \item $R_{r(i)} \sim U[0.75, 1.25]$ is a regional factor shared by buses with similar nominal voltage.
    \item $J_i \sim U[0.95, 1.05]$ is a local jitter factor.
\end{itemize}
Generator active power setpoints are scaled globally by factors sampled from $U[0.8, 1.2]$. Generators are modeled with sufficient reactive capability to maintain scheduled voltage setpoints.

To ensure the model learns generalized power flow physics rather than memorizing specific network impedances, we apply random jitter to the branch parameters. For every scenario, the electrical properties (resistance $R$, reactance $X$, and susceptance $B$) of all lines and transformers are independently scaled by a factor sampled from $U[0.9, 1.1]$ (i.e., $\pm 10\%$ variation).

Finally, to ensure robustness against topology changes, we generate $N\!-\!k$ contingencies by applying random branch outages. We consider up to two branch outages ($N\!-\!2$) when generating contingency scenarios.

\subsection{Graph Construction}
\noindent We construct a directed graph $\mathcal{G}=(\mathcal{V},\mathcal{E})$ where each bus is a node and each physical branch (line, transformer, series impedance, switch) is represented by reciprocal edges $(i\!\to\!j)$ and $(j\!\to\!i)$. This directed representation preserves direction-dependent branch parameters (Sec.~\ref{sec:node-edge-features}). Preprocessing yields node features $\mathbf{x}_i$ and edge features $\mathbf{e}_{ij}$. We also add a self-loop $(i\!\to\!i)$ for every node; its edge attributes carry diagonal network-admittance terms, while bus shunt admittance $(G_{\text{sh}}, B_{\text{sh}})$ is retained as node-level input.

\subsection{Node and Edge Features}\label{sec:node-edge-features}
\paragraph*{Node features} For each bus $i$, we provide a feature vector $\mathbf{x}_i \in \mathbb{R}^{23}$ that combines operational and topological information, together with a one-hot bus-type indicator:
\begin{itemize}
    \item \textbf{Nominal Values and Setpoints:} Nominal voltage ($V_n$), voltage setpoint ($V_m^{\text{set}}$), and active power setpoint ($P_g^{\text{set}}$).
    \item \textbf{Known Injections:} Active and reactive load ($P_{d}, Q_{d}$).
    \item \textbf{Physical Limits:} Voltage and reactive power limits ($V_m^{\min}, V_m^{\max}, Q_g^{\min}, Q_g^{\max}$).
    \item \textbf{Physics Parameters:} System base power ($S_{\text{base}}$), frequency ($f$), and bus shunt admittance ($G_{\text{sh}}, B_{\text{sh}}$).
    \item \textbf{Topological Features:} 1st- and 2nd-hop degree, weighted distance to slack, electrical betweenness, and aggregate neighbor net injection.
    \item \textbf{Neighbor Indicators:} Binary flags for 1st- and 2nd-hop neighbors being generators or slack buses.
\end{itemize}
The aggregate neighbor net injection feature is computed from generator setpoints and load inputs only, not from solved output targets.

\paragraph*{Edge features} Each directed branch edge $(i\!\to\!j)$ has a $7$-dimensional attribute vector obtained by concatenating series admittance ($Y^{\mathrm{ser}}_{ij}$), branch type ($\mathbf{t}_{ij}$), and thermal limit ($I^{\text{max}}_{ij}$):
\[
    \mathbf{e}_{ij}=\big[\ \underbrace{\Re(Y^{\mathrm{ser}}_{ij}),\ \Im(Y^{\mathrm{ser}}_{ij})}_{\text{admittance}} \;
        \underbrace{\mathbf{t}_{ij}}_{\text{type (one-hot)}\in\mathbb{R}^{4}} \;
        \underbrace{I^{\text{max}}_{ij}}_{\text{thermal limit}}\ \big] \in \mathbb{R}^{7}.
\]
Here, $\Re(\cdot)$ and $\Im(\cdot)$ are the real and imaginary parts. In implementation, $Y^{\mathrm{ser}}_{ij}$ is the directed off-diagonal branch-admittance edge term (including tap effects where applicable), so $\Re(Y^{\mathrm{ser}}_{ij})$ and $\Im(Y^{\mathrm{ser}}_{ij})$ are its conductance and susceptance components. $I^{\text{max}}_{ij}$ is a per-unit thermal-limit proxy computed as branch rating (MVA) normalized by system base power. For self-loops, the first two edge features use diagonal admittance terms, while type and rating features are set to zero. The nodal-balance terms $Y_{ik}$ in \eqref{eq:acpf-complex} are entries of the bus-admittance matrix $\mathbf{Y}$ assembled from network elements (including tap and shunt effects). Complete feature definitions are provided in Appendix~\ref{app:features}.

\subsection{Bus-Type Masks and Data Validation}
\noindent We define a binary training mask, $M \in \{0,1\}^{N\times 4}$, to identify the learning targets for each bus $i$ based on its type. A value of $1$ indicates an unknown target to be predicted. Table~\ref{tab:bus-masks} defines supervision over the model state vector $\mathbf{y}_i=[V_m,\delta,P_g,Q_g]$.

\paragraph*{Sampling balance}
We use a weighted random sampler with probability inversely proportional to the number of samples for each system to ensure balanced exposure across all systems (e.g., \code{case14} and \code{case\_illinois200}) during training.

\paragraph*{Feature and target scaling}
Node features are standardized (zero mean, unit variance) using statistics computed from the training set, and the same transformation is applied to the four target variables $(V_m, \delta, P_g, Q_g)$. Edge features are kept in physical units where appropriate. Predictions are inverse-transformed before evaluation and physics-loss computation, ensuring absolute-error metrics are reported in physical units, while normalized metrics are unitless.

\section{Model Architecture and Training}
\label{sec:gnn-model}
\begin{figure*}[t]
    \centering
    \includegraphics[width=0.95\linewidth]{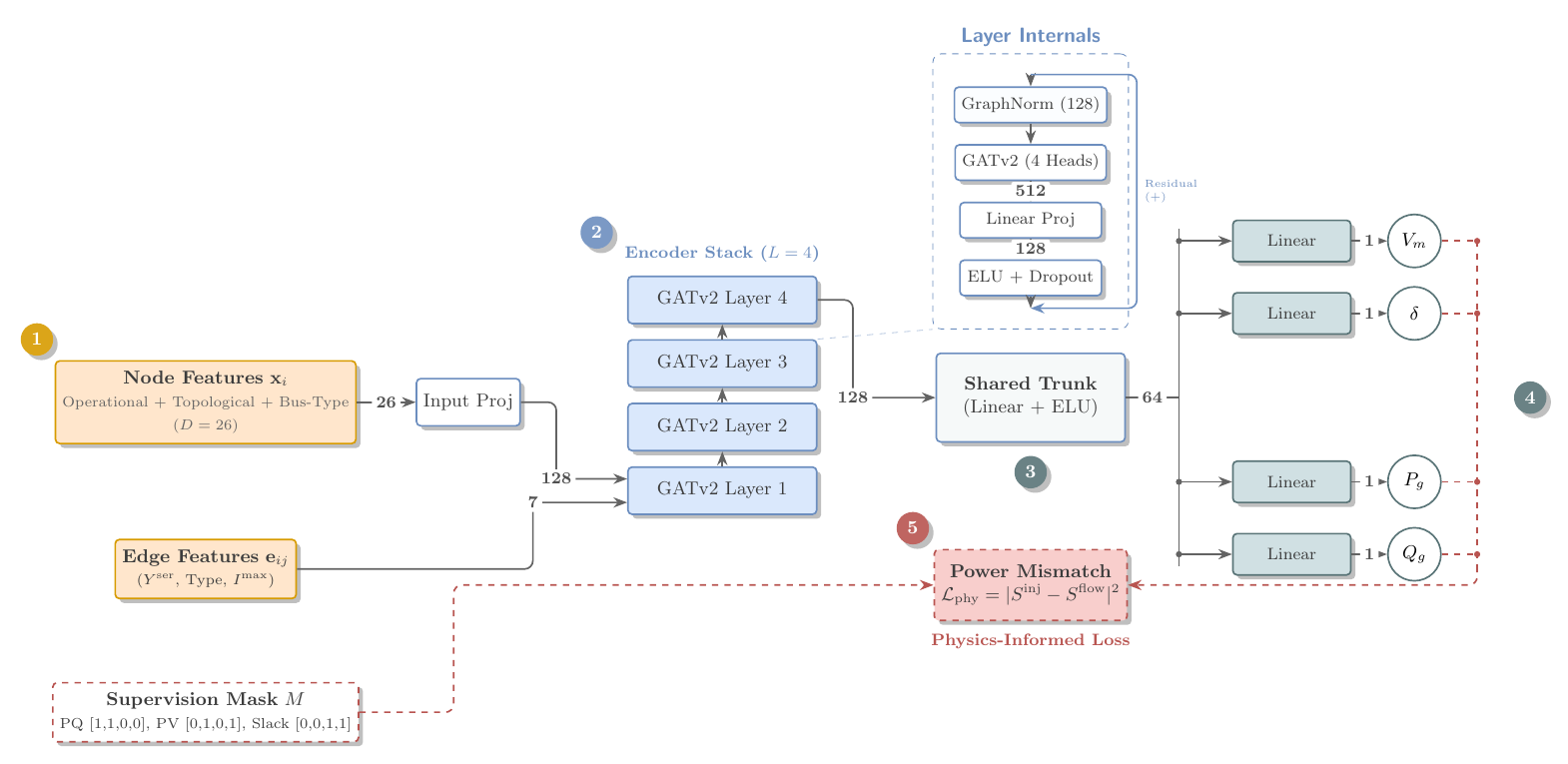}
    \caption{PowerModelsGAT-AI Architecture. The framework comprises (1) a bus-type-aware input stage with a supervision mask, (2) an encoder stack of 4 Pre-Norm Residual GATv2 blocks, (3) a shared decoding trunk, (4) multi-head outputs for $V_m,\,\delta,\,P_g,\,Q_g$, and (5) a physics-informed loss $\mathcal{L}_{\text{phy}}$. Red dashed lines indicate how predicted outputs and the supervision mask feed into the physics-informed loss $\mathcal{L}_{\text{phy}}$.}
    \label{fig:gnn_arch}
\end{figure*}
\noindent We design PMGAT-AI (Fig.~\ref{fig:gnn_arch}) as a deep GNN composed of stacked pre-norm residual blocks, followed by a multi-target output head.

\subsection{Pre-Norm Residual GNN Encoder}
\noindent The model maps input node features $\mathbf{x}_i$ to a hidden dimension $d_{\text{hidden}}$ using a linear projection. The resulting embeddings $\mathbf{h}^{(0)}$ are processed by $L$ pre-norm residual blocks (where normalization precedes each layer transformation).

For the intermediate layers $\ell \in \{1, \dots, L-1\}$, the update rule includes a non-linearity and dropout to deepen the representation:
\begin{align}
    \mathbf{h}^{(\text{norm})} & = \operatorname{GraphNorm}(\mathbf{h}^{(\ell-1)}, \mathbf{b}) \label{eq:prenorm}                                                                                           \\
    \mathbf{m}^{(\ell)}        & = \operatorname{GNN-Layer}(\mathbf{h}^{(\text{norm})}, \mathcal{E}, \mathbf{e}) \label{eq:gnn_layer}                                                                       \\
    \mathbf{h}^{(\ell)}        & = \mathbf{h}^{(\ell-1)} + \operatorname{Dropout}\!\left(\operatorname{ELU}\!\left(\mathbf{W}_{\text{head}}^{(\ell)}\mathbf{m}^{(\ell)}\right)\right) \label{eq:res_update}
\end{align}
where $\operatorname{GraphNorm}(\cdot, \mathbf{b})$ normalizes node features within each graph (identified by batch vector $\mathbf{b}$) with a learnable shift~\cite{cai2021graphnorm}, $\operatorname{GNN-Layer}$ is the GATv2 convolution defined in Sec.~\ref{sec:gatv2}, and $\mathbf{W}_{\text{head}}^{(\ell)}$ projects the concatenated attention heads back to $d_{\text{hidden}}$.

The final layer ($\ell=L$) prepares features for the readout head, aggregating attention heads by averaging rather than concatenation, and omitting the final activation:
\begin{equation}
    \mathbf{h}^{(L)} = \operatorname{Proj}_{\text{res}}(\mathbf{h}^{(L-1)}) + \operatorname{GNN-Layer}(\operatorname{GraphNorm}(\mathbf{h}^{(L-1)}), \mathcal{E}, \mathbf{e})
\end{equation}
where $\operatorname{Proj}_{\text{res}}$ aligns dimensions for the residual connection if necessary.

\subsection{GATv2 with Edge Features}
\label{sec:gatv2}
\noindent The $\operatorname{GNN-Layer}$ uses GATv2 \cite{brody2021attentive} with 7-dimensional edge features $\mathbf{e}_{ij}$ integrated into the edge-aware attention mechanism. For each of $H$ attention heads (indexed by $m$), the computation is:
\begin{align}
    \mathbf{z}_{ij}^{(m)} & = \mathbf{W}_t^{(m)}\mathbf{h}_i + \mathbf{W}_s^{(m)}\mathbf{h}_j + \mathbf{W}_e^{(m)}\mathbf{e}_{ij},\nonumber \\
    \tilde{e}_{ij}^{(m)}  & = \mathbf{a}^{(m)\top}\operatorname{LeakyReLU}\!\big(\mathbf{z}_{ij}^{(m)}\big),\nonumber                       \\
    \alpha_{ij}^{(m)}     & = \operatorname{softmax}_{j}\!\big(\tilde{e}_{ij}^{(m)}\big),\nonumber                                          \\
    \mathbf{m}_i^{(m)}    & = \sum_{j\in\mathcal{N}(i)}\alpha_{ij}^{(m)}\,\mathbf{W}_s^{(m)}\mathbf{h}_j.
    \label{eq:gatv2}
\end{align}
\paragraph*{Configuration} We set the hidden dimension $d_{\text{hidden}}{=}128$ and employ $L{=}4$ GATv2 layers with $H{=}4$ heads. Intermediate layer heads are concatenated, while the final layer heads are averaged. Physics-informed self-loops are explicitly included in the graph topology during construction (Sec.~\ref{sec:node-edge-features}) rather than added implicitly by the layer.

\subsection{Multi-Target Output Head}
\noindent A multi-head architecture maps the final node embedding $\mathbf{h}^{(L)}_i$ to the four-dimensional output vector $\hat{\mathbf{y}}_i$. A shared multilayer perceptron (MLP) trunk first extracts a latent representation $\mathbf{\xi}_i$:
\begin{equation}
    \mathbf{\xi}_i = \operatorname{ELU}(\mathbf{W}_{\text{trunk}}\mathbf{h}^{(L)}_i + \mathbf{b}_{\text{trunk}})
\end{equation}
This representation is projected by four separate linear layers to produce the state variable predictions:
\begin{align}
    \hat{V}_{m,i}    & = \mathbf{w}_{V_m}^\top \mathbf{\xi}_i + b_{V_m} \nonumber       \\
    \hat{\delta}_{i} & = \mathbf{w}_{\delta}^\top \mathbf{\xi}_i + b_{\delta} \nonumber \\
    \hat{P}_{g,i}    & = \mathbf{w}_{P_g}^\top \mathbf{\xi}_i + b_{P_g} \nonumber       \\
    \hat{Q}_{g,i}    & = \mathbf{w}_{Q_g}^\top \mathbf{\xi}_i + b_{Q_g}
\end{align}
\noindent This design uses a shared learned feature space while producing task-specific outputs.

\subsection{Loss Functions and Physics Constraints}
\label{sec:losses}
\noindent We formulate the training objective as a multi-task learning problem. Let the state at bus $i$ be $\mathbf{y}_i = [V_{m,i},\,\delta_i,\,P_{g,i},\,Q_{g,i}]$. We define five loss components $\mathcal{L}_\tau$ for $\tau \in \mathcal{C} = \{ V_m, \delta, P_g, Q_g, \text{phy} \}$, where $\text{phy}$ is the physics-mismatch penalty, which are dynamically balanced (Sec.~\ref{sec:multitask}).

A naive mean-squared error (MSE) over all buses implicitly biases training toward more frequent bus types (typically PQ buses) and larger systems. To reduce this bias from bus-type frequency and graph size, we compute the loss per graph using masks that include only unknown targets for each bus type (Table~\ref{tab:bus-masks}), then average across graphs.

Let $M_{i,\tau} \in \{0,1\}$ be the binary mask indicating if variable $\tau$ is an unknown target for bus $i$. For a batch of $N_{\mathcal{G}}$ graphs, the loss for state variable $\tau$ is computed by first normalizing the error within each graph $g$, and then averaging across graphs:
\begin{equation}
    \mathcal{L}_\tau = \frac{1}{N_{\mathcal{G}}} \sum_{g=1}^{N_{\mathcal{G}}} \left[ \frac{\sum_{i \in \mathcal{V}_g} M_{i,\tau} \cdot f_\tau(\hat{y}_{i,\tau}, y_{i,\tau})}{\sum_{i \in \mathcal{V}_g} M_{i,\tau} + \varepsilon} \right]
    \label{eq:masked-mse-unified}
\end{equation}
where $\mathcal{V}_g$ is the set of nodes in graph $g$ and $\varepsilon$ is a stability term. This formulation ensures that a small 14-bus system contributes equally to the gradient update as a 200-bus system, preventing bias toward larger systems.

\paragraph*{Node-wise error functions $f_\tau$}
For voltage magnitudes and power injections ($\tau \in \{V_m, P_g, Q_g\}$), $f_\tau$ is the standard squared error on standardized targets:
\begin{equation}
    f_{\tau}(\hat{y}, y) = (\hat{y} - y)^2
\end{equation}

For voltage angles, standard MSE is not suitable because angles are periodic. To address this, we optimize the wrapped residual in degrees; in our experiments, this improved convergence stability and final predictive performance relative to a radian-domain loss. Let $\Delta_i=\hat{\delta}_i^{(\mathrm{rad})}-\delta_i^{(\mathrm{rad})}$ be the angular difference in radians. The wrapped loss is
\begin{equation}
    f_{\delta}(\hat{\delta}_i, \delta_i)
    = \left(\frac{180}{\pi}\,\operatorname{atan2}\!\left(\sin\Delta_i,\cos\Delta_i\right)\right)^2
\end{equation}
where $\operatorname{atan2}(\sin\Delta,\cos\Delta)$ returns the principal value in $(-\pi,\pi]$, so the error reflects the shortest arc.

\paragraph*{Physics-Informed Power Mismatch}
\noindent We enforce physical consistency by minimizing bus-level power-mismatch errors derived from the nodal power balance in \eqref{eq:acpf-complex}. Specifically, we penalize the deviation between the hybrid net injection $S^{\mathrm{inj,hyb}}_i$ and the calculated flow $\hat{S}^{\mathrm{flow}}_i$. To prevent information leakage, we define $S^{\mathrm{inj,hyb}}_i$ by mixing model predictions with reference data according to the bus mask:
\begin{equation}
    S^{\mathrm{inj,hyb}}_i = (P_{g,i}^{\text{hyb}} - P_{d,i}) + j(Q_{g,i}^{\text{hyb}} - Q_{d,i})
\end{equation}
where the hybrid generation terms combine the predicted ($\hat{P}_g, \hat{Q}_g$) and known ($P_g, Q_g$) values:
\begin{align}
    P_{g,i}^{\text{hyb}} & = M_{i,P_g} \hat{P}_{g,i} + (1 - M_{i,P_g}) P_{g,i}, \\
    Q_{g,i}^{\text{hyb}} & = M_{i,Q_g} \hat{Q}_{g,i} + (1 - M_{i,Q_g}) Q_{g,i}.
\end{align}
Here, $P_{d,i}$ and $Q_{d,i}$ are ZIP-equivalent loads evaluated at $\hat{V}_{m,i}$.

The calculated flow is the nodal power balance \eqref{eq:acpf-complex} evaluated at predicted voltages, where $\hat{V}_i = \hat{V}_{m,i} e^{j\hat{\delta}_i}$ is the complex voltage constructed from the predicted magnitude and angle:
\begin{equation}
    \hat{S}^{\mathrm{flow}}_i = \hat{V}_i \sum_{k \in \mathcal{N}(i)} Y_{ik}^* \hat{V}_k^*
\end{equation}
The physics loss is the mean squared modulus of the mismatch, computed in \SI{64}{\bit} precision for numerical stability and averaged per graph as in \eqref{eq:masked-mse-unified}:
\begin{equation}
    \mathcal{L}_{\text{phy}} = \frac{1}{N_{\mathcal{G}}}\sum_{g=1}^{N_{\mathcal{G}}} \left[ \frac{1}{|\mathcal{V}_g|} \sum_{i \in \mathcal{V}_g} \big| S^{\mathrm{inj,hyb}}_i - \hat{S}^{\mathrm{flow}}_i \big|^2 \right]
\end{equation}

\subsection{Multi-Task Optimization}
\label{sec:multitask}
\noindent The training objective involves minimizing five distinct loss components with different scales and optimization difficulty. Naive summation with fixed weights requires extensive hyperparameter tuning and can cause one loss to dominate the gradient updates over the others.

To address this, we treat the problem as a multi-task learning challenge and employ homoscedastic uncertainty weighting~\cite{kendall2018multi}. This method interprets the relative weight of each task as a learnable uncertainty parameter $\sigma_\tau$, derived from maximizing the Gaussian likelihood of the multi-task objective.

Let $\mathcal{C} = \{ V_m, \delta, P_g, Q_g, \text{phy} \}$ be the set of task indices defined in Sec.~\ref{sec:losses}, comprising the four supervised state variables and the physics mismatch. The unified training objective $\mathcal{L}_{\text{total}}(\bm{\theta}, \bm{\sigma})$ is:
\begin{equation}
    \mathcal{L}_{\text{total}}(\bm{\theta}, \bm{\sigma}) = \sum_{\tau \in \mathcal{C}} \frac{1}{2\sigma_\tau^2} \mathcal{L}_\tau(\bm{\theta}) + \sum_{\tau \in \mathcal{C}} \log \sigma_\tau
    \label{eq:uncertainty}
\end{equation}
where $\bm{\theta}$ represents the model parameters and $\bm{\sigma} = \{\sigma_\tau\}_{\tau \in \mathcal{C}}$ are learnable noise scalars optimized simultaneously with the model parameters. The term $\log \sigma_\tau$ acts as a regularizer to prevent the trivial solution $\sigma_\tau \to \infty$. This approach dynamically balances the supervised losses and the physics-informed constraint without manual tuning, with learned weights that adapt to reflect the relative difficulty of each prediction task.
In implementation, uncertainty weighting is activated after an initial static-loss warmup period (50 epochs).

\subsection{Optimization and Early Stopping}
\label{sec:training}

\noindent We minimize \eqref{eq:uncertainty} using the AdamW optimizer with an initial learning rate $\eta_0 = 10^{-3}$ and weight decay $10^{-5}$. The learning rate follows a schedule comprising a linear warmup for the first $E_w$ epochs, followed by cosine annealing decay~\cite{loshchilovstochastic} to $\eta_{\min} = 10^{-5}$ for the remainder of the training budget $E$.

We trained on an NVIDIA RTX 4000 Ada GPU (\SI{20}{\giga\byte} VRAM). We used a baseline batch size of 128, reducing this value for large-scale systems and during multi-system learning to accommodate memory constraints.
The implementation is based on PyTorch and PyTorch Geometric, with NumPy, SciPy, NetworkX, Matplotlib, and scikit-learn for data processing, graph utilities, visualization, and evaluation~\cite{paszke2019pytorch,fey2025pyg,harris2020array,scipy2020nmeth,SciPyProceedings_11,Hunter:2007,scikit-learn}.

We use a two-phase early stopping strategy. After the main validation metric plateaus, we keep a checkpoint only if it reduces the physics mismatch $\mathcal{L}_{\text{phy}}$ and does not reduce overall performance by more than \SI{5}{\percent}. This gives a final model that stays physically consistent while maintaining low supervised loss.

\section{Continual Learning for On-the-Fly Adaptation}
\label{sec:continual}
\noindent Operational power grids are dynamic; topological changes (switching, outages) or extreme loading conditions can effectively alter the data distribution. To adapt to new scenarios and new systems without full retraining, we combine experience replay and EWC within a continual-learning framework to mitigate catastrophic forgetting.

\paragraph*{Experience Replay}
To enable the model to learn from both old and new scenarios or systems, we employ experience replay. We maintain a replay buffer $\mathcal{D}_{\text{rep}} \subset \mathcal{D}_{\text{old}}$ of scenarios from the systems on which the model was originally trained, sampled uniformly at random.

During adaptation, the new data $\mathcal{D}_{\text{new}}$ may consist of stressed scenarios from existing systems or scenarios from a new system. Because shifts are larger when adapting across structurally different systems, we mix $\mathcal{D}_{\text{new}}$ with replay data $\mathcal{D}_{\text{rep}}$ to reduce forgetting while learning the new target systems. We construct a mixed training set $\mathcal{D}_{\text{mix}} = \mathcal{D}_{\text{new}} \cup \mathcal{D}_{\text{rep}}$, ensuring that the model retains performance on base systems while adapting to the new systems or operating conditions.

\paragraph*{Elastic Weight Consolidation (EWC)}
While experience replay helps mitigate data distribution shifts, it does not prevent important model parameters from changing. To mitigate this, we augment the loss with EWC~\cite{kirkpatrick2017overcoming}, which constrains parameters that are critical for performance on base systems.

Let $\bm{\theta}^*$ be the parameters learned during initial training. We compute the diagonal Fisher Information Matrix $F$, where each element $F_j$ approximates the importance of parameter $\theta_j$ for maintaining performance on base systems. The regularization penalty is:
\begin{equation}
    \mathcal{L}_{\text{EWC}}(\bm{\theta}) = \lambda_{\text{ewc}} \sum_{j} F_j (\theta_j - \theta^*_j)^2
    \label{eq:ewc}
\end{equation}
This regularization constrains parameters with high $F_j$ (critical for power flow solutions on base systems), while allowing low-sensitivity parameters to adapt to new systems or operating conditions.

The final on-the-fly objective unifies the learning signal from the mixed data batch $\mathcal{B} \sim \mathcal{D}_{\text{mix}}$ with the EWC penalty:
\begin{equation}
    \mathcal{L}_{\text{online}} = \mathcal{L}_{\text{total}}(\bm{\theta}, \bm{\sigma}, \mathcal{B}) + \mathcal{L}_{\text{EWC}}(\bm{\theta})
    \label{eq:online}
\end{equation}
where $\mathcal{L}_{\text{total}}$ is the multi-task objective from \eqref{eq:uncertainty}.

\section{Evaluation Setup}
\label{sec:setup}

\subsection{Dataset and Evaluation Regimes}
PMGAT-AI is evaluated on 14 benchmark systems ranging from small-scale systems (e.g., \code{case4gs}, \code{case14}) to large-scale systems (e.g., \code{case1354pegase}, \code{case6470rte}). We define two distinct evaluation regimes:

\subsubsection{Regime 1: System-Specific Baselines}
We establish baselines by training independent PMGAT-AI models for each of the 14 systems, using 10,000 scenarios per system. We evaluate two operating regimes: (1)~\emph{Normal operation} ($N\!-\!0$), where models are trained and tested on randomized load and generation profiles with fixed topology; and (2)~\emph{Contingency} ($N\!-\!2$), where models are retrained on datasets with randomized load and generation profiles and up to two random branch outages per scenario to improve robustness (Appendix~\ref{app:contingency-stats}, Table~\ref{tab:contingency-specialized}).

\subsubsection{Regime 2: Multi-System and Continual Learning}
This regime evaluates multi-system learning under $N\!-\!2$ contingency conditions. Because batching multiple large graphs is memory-intensive, \code{case6470rte} is excluded from unified training. We sample 2,000 scenarios per system, resulting in 26,000 scenarios for the unified model trained on 13 systems and 24,000 for the base model trained on 12 systems. We first train the unified model on all 13 systems (Table~\ref{tab:unified-n2}). This model is also used for interpretability analysis in Sec.~\ref{sec:interpretability}. Then, we evaluate continual learning by pre-training a base model on 12 systems (excluding \code{case1354pegase}) and adapting it to \code{case1354pegase} using the EWC+Replay strategy in Sec.~\ref{sec:continual}. Detailed contingency statistics are provided in Appendix~\ref{app:contingency-stats}, Table~\ref{tab:contingency-unified}.

For very small systems (notably \code{case4gs} and \code{case9}), feasible $N\!-\!2$ samples are absent due to islanding or failure of AC power flow to converge; these systems contribute only $N\!-\!0$ and $N\!-\!1$ samples.

\subsection{Evaluation Protocol}
We randomly split all the datasets into training (80\%), validation (10\%), and test (10\%) sets, stratified by system, then evaluate all models on the held-out test sets. Metrics are computed only for \emph{unknown} state variables as defined by each bus type (Table~\ref{tab:bus-masks}). This prevents known fixed setpoints from artificially inflating performance metrics.

We use the following metrics to evaluate different aspects of model performance:
\begin{itemize}
    \item \textbf{Absolute and Squared Error:} We report root mean squared error (RMSE) and mean absolute error (MAE). For voltage angles, errors are wrapped to $[-180^{\circ}, 180^{\circ}]$ before computation.
    \item \textbf{Scale-Normalized Error:} We report normalized mean absolute error (NMAE), defined as MAE divided by the range of target values for each target.
    \item \textbf{Coefficient of Determination ($R^2$):} We report $R^2$. For voltage angles, $R^2$ is computed with wrapped residuals and a circular-mean reference so that it is consistent with the angle MAE and RMSE definitions.
    \item \textbf{Knowledge Loss:} We quantify forgetting with Knowledge Loss ($\mathcal{K}$). For state variable $\tau \in \{V_m,\delta,P_g,Q_g\}$, error metric $\mu$ (e.g., NMAE, MAE), and previously learned system $s$:
          \begin{equation}
              \mathcal{K}_{\tau}^{\mu}(s) = \mathrm{Err}_{\tau,\mu}^{\text{post}}(s) - \mathrm{Err}_{\tau,\mu}^{\text{pre}}(s)
          \end{equation}
          where $\mathrm{Err}^{\text{pre}}$ and $\mathrm{Err}^{\text{post}}$ are errors before and after fine-tuning on the new task. Positive $\mathcal{K}$ indicates forgetting; negative $\mathcal{K}$ indicates improved post-adaptation performance. Across base systems $\mathcal{S}$, we report:
          \begin{equation}
              \bar{\mathcal{K}}_{\tau}^{\mu} = \frac{1}{|\mathcal{S}|}\sum_{s \in \mathcal{S}} \mathcal{K}_{\tau}^{\mu}(s)
          \end{equation}
\end{itemize}

\section{Results}
\label{sec:results}

Results are organized by the two regimes defined in Sec.~\ref{sec:setup}.

\subsection{Regime 1: Baseline System Performance ($N\!-\!0$)}
\label{subsec:n0-results}

The detailed per-system metrics in Appendix~\ref{app:baselines}, Table~\ref{tab:single-topo-n0} establish baseline performance under normal $N\!-\!0$ conditions.

For voltage magnitudes ($V_m$), MAE on standard IEEE benchmark systems (e.g., \code{case14}, \code{case30}) is on the order of $10^{-4}~\si{\pu}$, corresponding to an NMAE of approximately $0.2\%$--$0.3\%$. For the larger \code{case6470rte} system, NMAE is approximately $0.98\%$, and $R^2$ remains above $0.94$, showing that the model captures system-level voltage variation.

We observe that voltage angle ($\delta$) error increases with system size (MAE $\approx 0.05^\circ$ for \code{case14} versus $\approx 5.0^\circ$ for \code{case6470rte}). A likely contributor is reference-angle drift in larger networks~\cite{caro2016uncertainty}. Since nodal power balance depends on angle differences between connected buses (\eqref{eq:p_line}, \eqref{eq:q_line}), reference-angle drift can increase absolute angle MAE without degrading the inter-bus differences that govern power flow. Even so, angle $R^2$ remains high ($\approx 0.99$ for \code{case6470rte}).

Active ($P_g$) and reactive ($Q_g$) predictions at slack and PV buses also show low error, with $P_g$ $R^2$ consistently above 0.99 across systems. Overall, these metrics show that the model preserves physically consistent power flow relationships across the evaluated systems and scenarios.

\subsection{Regime 1: Robustness to Contingencies ($N\!-\!2$)}
\label{subsec:n2-results}

To assess robustness to topology changes within a system, Regime 1 also evaluates each system under $N\!-\!2$ contingencies, where random branch outages redistribute branch flows and bus-voltage profiles, creating many topology combinations and a harder prediction task. Complete per-system results are provided in Appendix~\ref{app:baselines}, Table~\ref{tab:single-topo-n2}.

Compared with the $N\!-\!0$ baseline, we observe marginal error increases (e.g., in \code{case14}, voltage-magnitude MAE increases from $3.3\times10^{-4}$ to $7.2\times10^{-4}~\si{\pu}$), but errors remain small. Voltage NMAE stays below $0.5\%$ for most systems, and angle $R^2$ stays near $0.99$. Even in the challenging \code{case6470rte} system under $N\!-\!2$ contingency, angle $R^2$ is $\approx 0.99$, close to its $N\!-\!0$ value. This is consistent with the GATv2 dynamic attention mechanism reweighting branch importance as connectivity, loading, and branch attributes change.

\subsection{Regime 2: Multi-System and Continual Learning}
\label{subsec:unified-results}

We evaluate PMGAT-AI across multiple trained systems and for adaptation to new systems. We first assess a unified model trained on 13 systems, then evaluate continual adaptation by fine-tuning a base model trained on 12 systems to \code{case1354pegase}.

\subsubsection{Unified 13-System Performance}
We analyze the unified model trained on 13 systems. Detailed results are provided in Appendix~\ref{app:unified}, Tables~\ref{tab:unified-n2} and~\ref{tab:unified-n0-n1-n2}.

Compared with system-specific models, the unified model can be used across multiple systems, with a small increase in the per-system error. For example, on \code{case14}, voltage NMAE increases from about 0.15\% for the system-specific model (Regime 1) to about 0.76\% for the unified model.

\begin{figure*}[t]
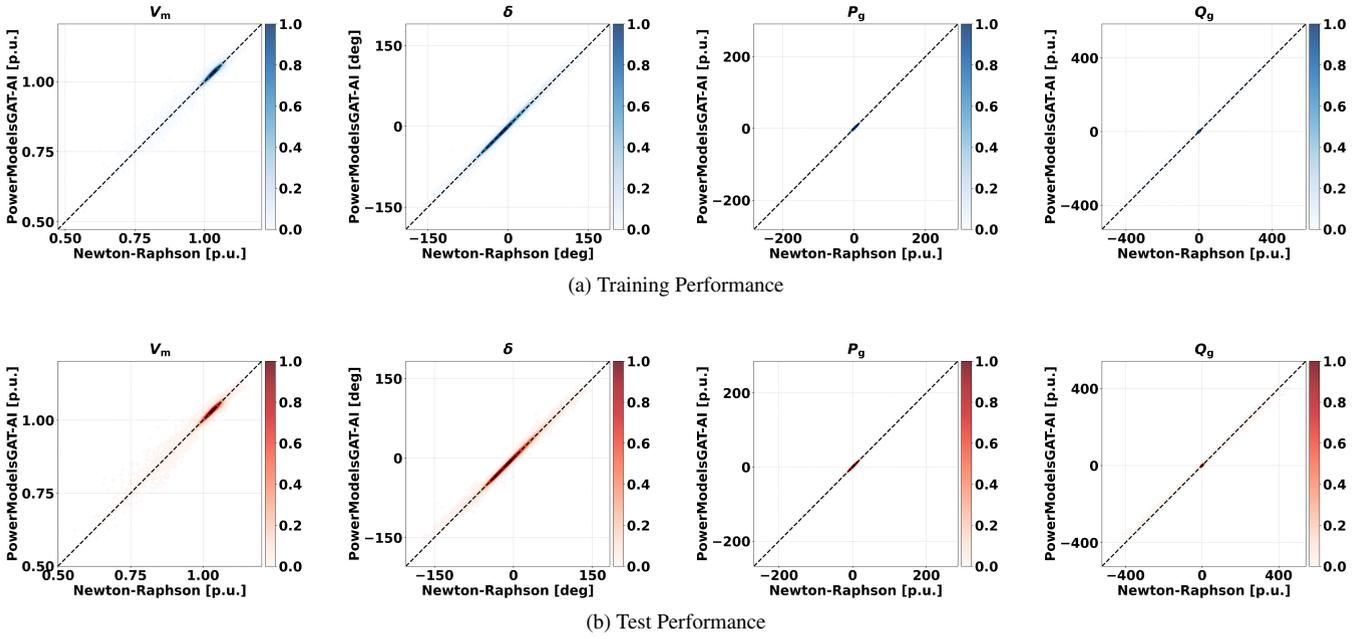

    \centering
    \subfloat[Training Performance]{\includegraphics[width=\linewidth]{images/aggregated_train_unknowns.png}\label{fig:agg_train}}
    \\
    \vspace{0.5em}
    \subfloat[Test Performance]{\includegraphics[width=\linewidth]{images/aggregated_test_unknowns.png}\label{fig:agg_test}}
    \caption{PowerModelsGAT-AI Performance across 13 Power Systems. Parity plots compare model predictions (y-axis) against the Newton--Raphson reference solution (x-axis) for all unknown variables in the unified 13-system training and test sets; color intensity represents point density.}
    \label{fig:agg_perf}
\end{figure*}

Across the 13 trained systems, the average voltage NMAE is $0.89\%$, and active-power $R^2$ exceeds $0.97$ (Fig.~\ref{fig:agg_perf}), with Fig.~\ref{fig:agg_train} and Fig.~\ref{fig:agg_test} showing similar prediction versus reference trends across all predicted variables. Per-system plots are provided in the Supplementary Material (Figs.~S1--S13).

At the bus level, prediction error is typically lowest in small-scale systems such as \code{case4gs} and \code{case30}, where predictions closely match the reference values. In larger systems such as \code{case1354pegase}, target variables (especially $\delta$, $P_g$, and $Q_g$) span wider ranges across buses and scenarios. As a result, absolute errors can be larger even when NMAE remains relatively low. Branch outages further widen these ranges and increase operating stress, which increases angle and power-injection deviations. Additional higher-error examples are provided in the Supplementary Material (Fig.~S14).

\subsubsection{Continual Learning: Adaptation Without Forgetting}
While unified training works well, in practice new systems may need to be incorporated without full retraining. Prior work highlights persistent generalization and scalability challenges under changes in system size and topology~\cite{pineda2025beyond,khaloie2025review}. To quantify this, we evaluated a base model trained on 12 systems against the held-out \code{case1354pegase} system. As shown in Table~\ref{tab:target-performance}, the unadapted base model performs poorly on the target system, with voltage NMAE of $12.98\%$, angle NMAE of $11.96\%$, and negative $R^2$ values meaning predictions worse than the mean. These gaps make clear that cross-system generalization is not automatic and requires explicit adaptation strategies.

We evaluated all models on identical test sets and used $\lambda_{\text{ewc}}=0.5$ with a replay ratio of 0.3 to balance adaptation and forgetting. We compared this setting against a naive baseline ($\lambda_{\text{ewc}}=0$, no replay).

\begin{table}[t]
    \centering
    \caption{Target System (\code{case1354pegase}) Performance After Fine-Tuning.}
    \label{tab:target-performance}
    \renewcommand{\arraystretch}{1.15}
    \footnotesize
    \begin{tabular*}{\columnwidth}{@{\extracolsep{\fill}} l c c c c c c @{}}
        \toprule
        & \multicolumn{2}{c}{\textbf{Base (Unadapted)}} & \multicolumn{2}{c}{\textbf{Naive}} & \multicolumn{2}{c}{\textbf{EWC+Replay}}                                                     \\
        \cmidrule(lr){2-3} \cmidrule(lr){4-5} \cmidrule(l){6-7}
        \textbf{Target} & \textbf{NMAE\%}                               & \textbf{R\textsuperscript{2}}      & \textbf{NMAE\%}                         & \textbf{R\textsuperscript{2}} & \textbf{NMAE\%} & \textbf{R\textsuperscript{2}} \\
        \midrule
        $V_m$           & 12.98                                         & \rTwoDisplay{-2.36}               & \textbf{1.44}                           & \rTwoDisplay{0.945} & 1.90            & \rTwoDisplay{0.907} \\
        $\delta$        & 11.96                                         & \rTwoDisplay{-0.01}               & \textbf{0.61}                           & \rTwoDisplay{0.997} & 0.68            & \rTwoDisplay{0.997} \\
        $P_g$           & 20.79                                         & \rTwoDisplay{0.01}                & \textbf{0.42}                           & \rTwoDisplay{0.999} & 0.45            & \rTwoDisplay{0.999} \\
        $Q_g$           & 1.61                                          & \rTwoDisplay{0.02}                & \textbf{0.15}                           & \rTwoDisplay{0.994} & 0.17            & \rTwoDisplay{0.994} \\
        \bottomrule
    \end{tabular*}
    \vspace{0.3em}
    \newline
    {\footnotesize Target-system performance is shown here; base-system forgetting is summarized in Table~\ref{tab:continual-learning}.}
\end{table}

\begin{table}[t]
    \centering
    \caption{Average Knowledge Loss ($\bar{\mathcal{K}}_{\tau}^{\text{NMAE}}$) on Base Systems.}
    \label{tab:continual-learning}
    \renewcommand{\arraystretch}{1.15}
    \footnotesize
    \begin{tabular*}{\columnwidth}{@{\extracolsep{\fill}} l c c @{}}
        \toprule
        \textbf{State Variable}   & \textbf{Naive} & \textbf{EWC+Replay} \\
        \midrule
        Voltage Magnitude ($V_m$) & 12.00          & \textbf{0.26}       \\
        Voltage Angle ($\delta$)  & 138.23         & \textbf{0.12}       \\
        Active Power ($P_g$)      & 1891.82        & \textbf{1.25}       \\
        Reactive Power ($Q_g$)    & 103.79         & \textbf{1.54}       \\
        \bottomrule
    \end{tabular*}
    \vspace{0.3em}
    \newline
    {\footnotesize Values are in percentage points of NMAE change ($\bar{\mathcal{K}}_{\tau}^{\text{NMAE}}$), averaged across 12 base systems. Lower is better; $\bar{\mathcal{K}} \approx 0$ indicates minimal forgetting on base systems.}
\end{table}

On \code{case1354pegase}, both strategies reach $R^2 > 0.99$ for voltage angles and power injections (Table~\ref{tab:target-performance}). Naive fine-tuning gives slightly better target $V_m$ NMAE ($1.44\%$ versus $1.90\%$), but it causes severe forgetting on base systems ($\bar{\mathcal{K}}_{P_g}^{\text{NMAE}} \approx 1892\%$, $\bar{\mathcal{K}}_{\delta}^{\text{NMAE}} \approx 138\%$; Table~\ref{tab:continual-learning}). Naive fine-tuning is not suitable for long-term cross-system adaptation.

Our EWC+Replay strategy addresses this problem. Although target performance is slightly lower than naive fine-tuning, it maintains low knowledge loss across all state variables ($\bar{\mathcal{K}}_{V_m}^{\text{NMAE}} = 0.26\%$, $\bar{\mathcal{K}}_{\delta}^{\text{NMAE}} = 0.12\%$, $\bar{\mathcal{K}}_{P_g}^{\text{NMAE}} = 1.25\%$, $\bar{\mathcal{K}}_{Q_g}^{\text{NMAE}} = 1.54\%$). The same trend appears with MAE ($\bar{\mathcal{K}}_{\delta}^{\text{MAE}} = 0.16^{\circ}$ versus $74^{\circ}$ for naive; see Appendix~\ref{app:continual}).

Taken together, these results confirm that the model can adapt to new systems while minimizing forgetting on base systems. Visual summaries are provided in Supplementary Material (Figs.~S15 and S16).

\section{Interpretability Analysis}
\label{sec:interpretability}

To assess PMGAT-AI's ability to capture established physical relationships from learned representations rather than relying on spurious statistical correlations, we analyze the model from two perspectives: \textit{branch importance}, quantified using learned attention weights, and \textit{bus feature sensitivity}, assessed using integrated gradients. Attention weights are useful for interpretation, but they do not by themselves establish causal influence~\cite{jain-wallace-2019-attention,wiegreffe-pinter-2019-attention}. However, consistent correlations with known electrical parameters provide supporting evidence of alignment with electrical relationships.

\subsection{Identification of Critical Topological Structures}
\begin{figure*}[t]
    \centering
    \subfloat[\code{case14}]{
        \includegraphics[width=0.48\linewidth]{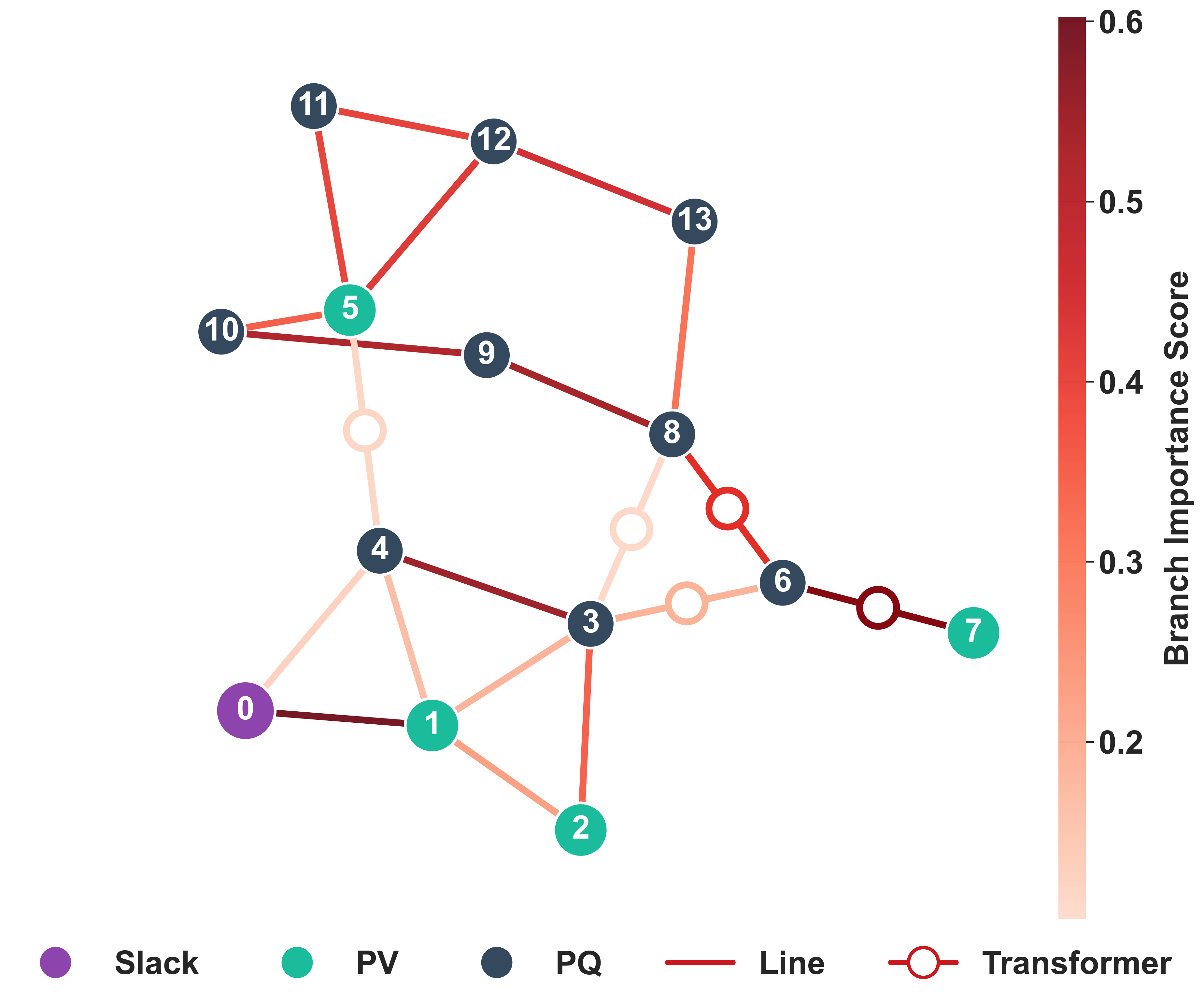}
        \label{fig:attn_case14}
    }
    \hfill
    \subfloat[\code{case57}]{
        \includegraphics[width=0.48\linewidth]{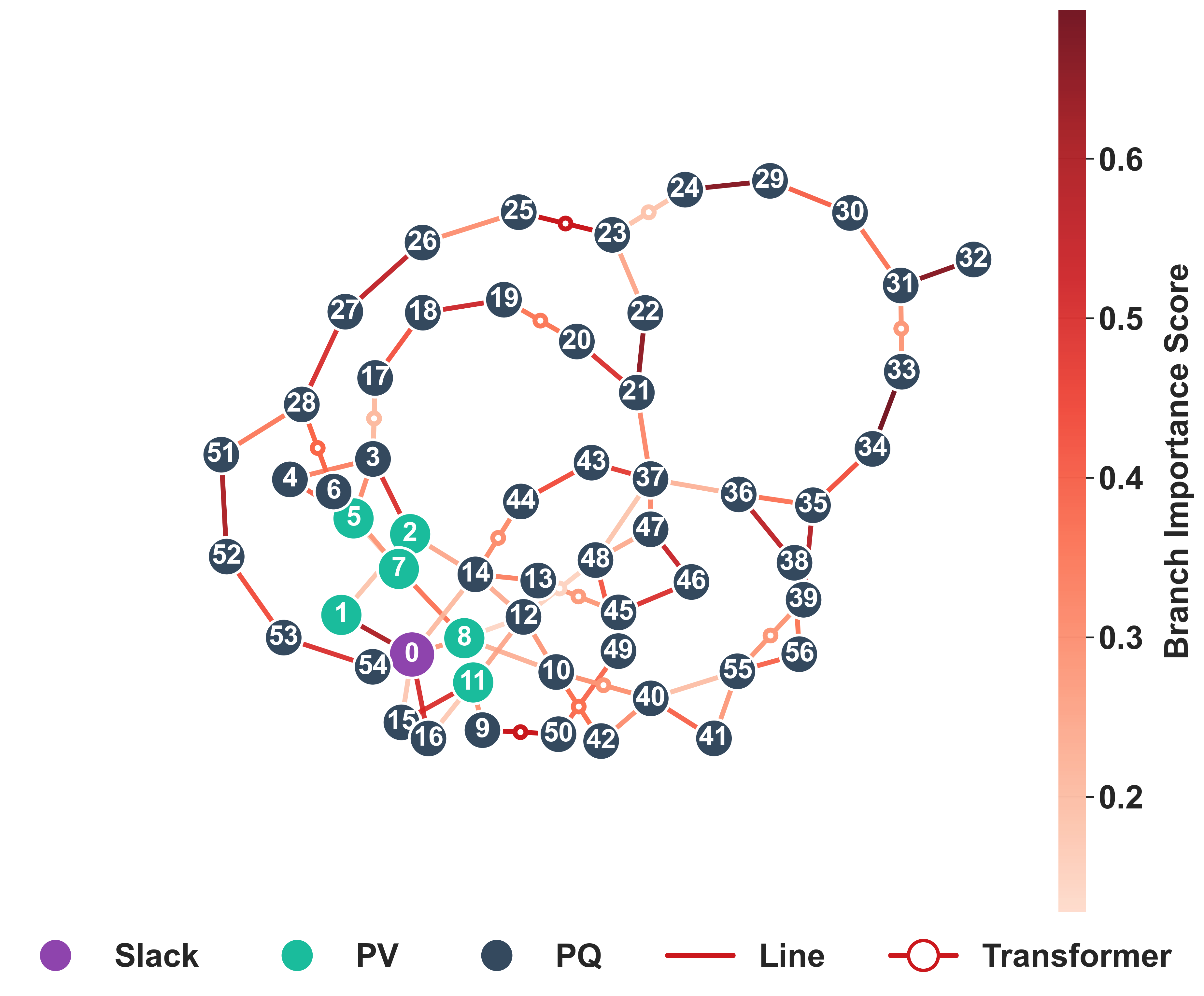}
        \label{fig:attn_case57}
    }
    \caption{Learned Branch Importance Maps. Darker edges indicate higher importance. (a) \code{case14}: the model assigns higher importance to key branches. (b) \code{case57}: importance scores vary across the system.}
    \label{fig:attn_maps}
\end{figure*}

To quantify topological influence, we use the learned attention coefficients from the final GATv2 layer. For each branch $(i,j)$, we define a \emph{branch importance score} as the mean attention weight
\begin{equation}
    \bar{\alpha}_{ij} = \frac{1}{|\mathcal{D}|}\sum_{c \in \mathcal{D}} \left( \frac{1}{H}\sum_{m=1}^{H}\alpha_{ij}^{(m, c)} \right),
\end{equation}
averaged across all $H$ attention heads and all test scenarios $c \in \mathcal{D}$. We use these scores for topology visualization (Fig.~\ref{fig:attn_maps}). For correlation analysis (Fig.~\ref{fig:combined_analysis}a), we compute feature--attention correlations from branch-level attention within each system and report the equal-weight average across systems.

Fig.~\ref{fig:attn_maps} illustrates the learned importance scores for two representative systems, \code{case14} (Fig.~\ref{fig:attn_case14}) and \code{case57} (Fig.~\ref{fig:attn_case57}). The model does not attend uniformly to all branches. Instead, it assigns higher importance to a subset of key branches. In \code{case14}, the model assigns higher importance to key branches (dark red). Transformer branches show mixed importance: several receive low scores, whereas others are among the most influential connections. The model responds to system-specific coupling rather than simply down-weighting transformers.

\subsection{Physical Drivers of Branch Importance}

\begin{figure*}[t]
    \centering
    \subfloat[Feature Correlation]{
        \includegraphics[width=0.48\linewidth]{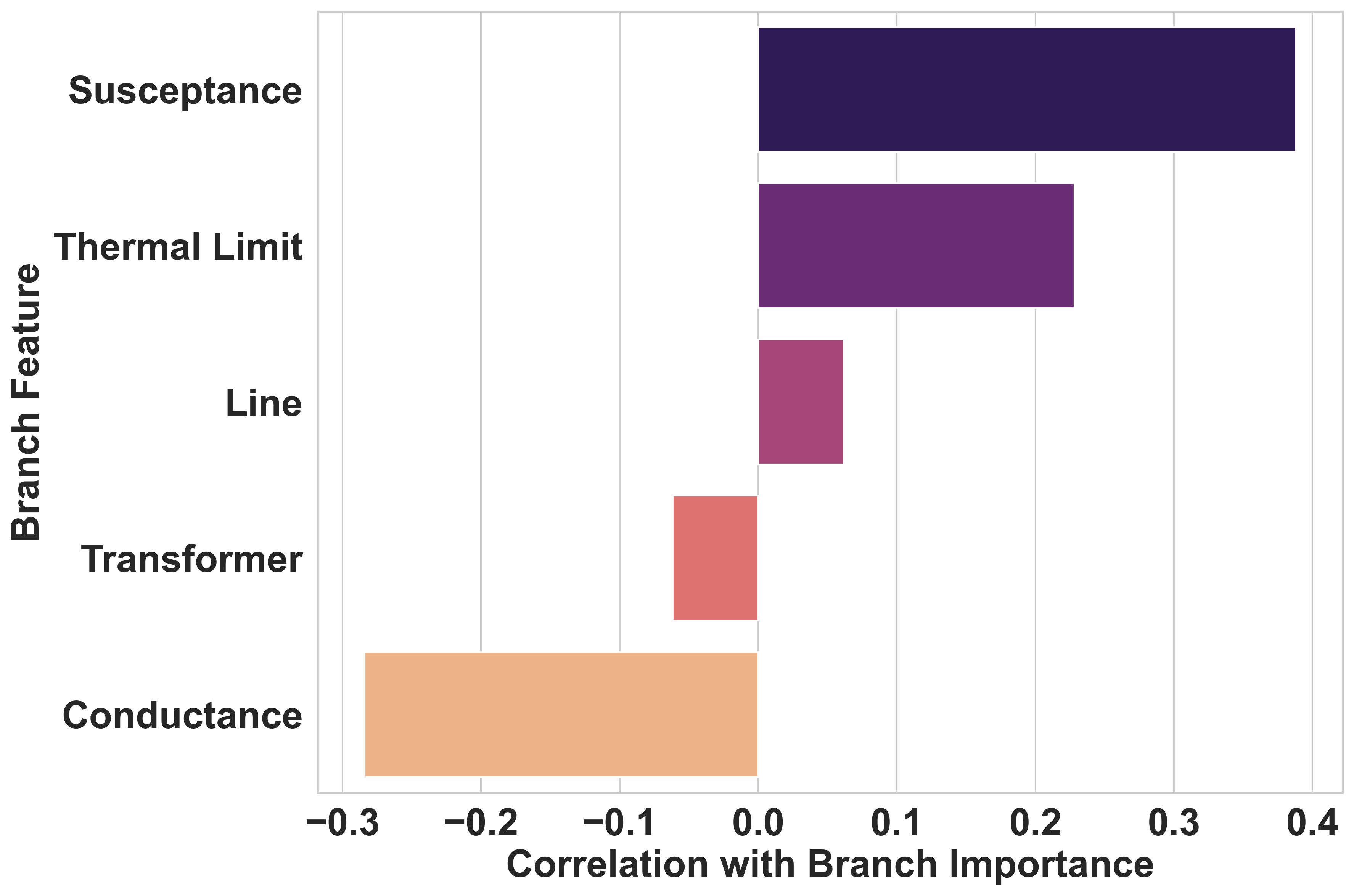}
        \label{fig:global_corr}
    }
    \hfill
    \subfloat[Distribution by System]{
        \includegraphics[width=0.48\linewidth]{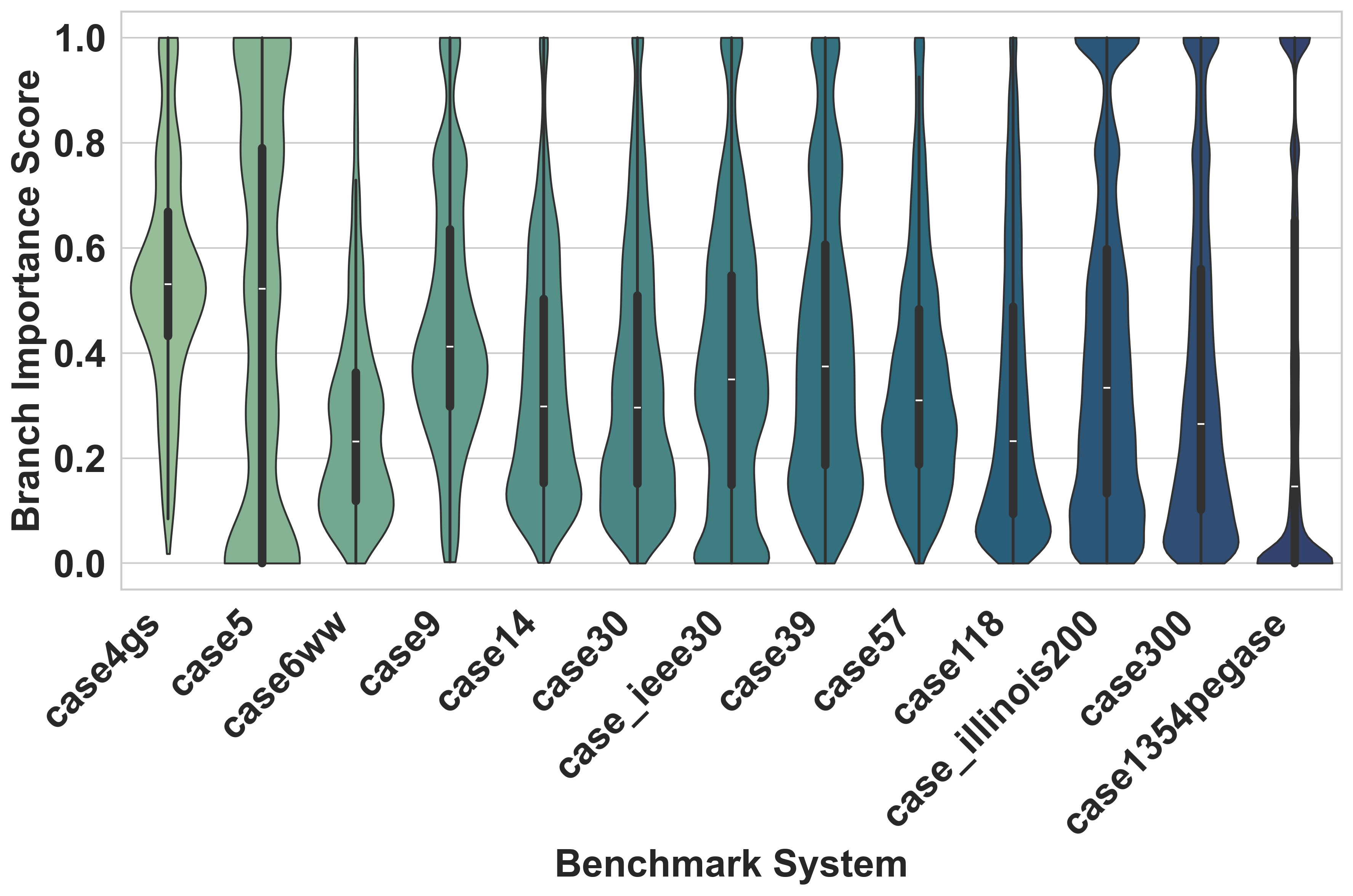}
        \label{fig:dist}
    }
    \caption{Analysis of Branch Importance Scores. (a) Correlation between branch importance and physical branch parameters. (b) Distribution of branch importance scores across benchmark systems.}
    \label{fig:combined_analysis}
\end{figure*}

To interpret why some branches receive higher importance, we compute Pearson correlations ($r$) between branch-importance scores and raw electrical branch parameters (Fig.~\ref{fig:combined_analysis}a). Correlations are computed within each system and then averaged with equal system weight.

Three patterns hold across systems:
\begin{enumerate}
    \item \textbf{Susceptance ($\Im(Y^{\mathrm{ser}})$):} Susceptance has the strongest positive correlation ($r \approx 0.38$). This reflects AC power flow relationships, where real-power flow is strongly tied to branch reactance and susceptance and shared among parallel branches according to their impedances~\cite{stott1974fast,grainger1994power}.
    \item \textbf{Thermal Limit ($I^{\text{max}}$):} Thermal limit is also positively correlated ($r \approx 0.22$). Although thermal rating is not a direct admittance term, higher-rated branches are often electrically influential, so this association is physically consistent. The effect is system-dependent, and in some systems with near-constant branch ratings the thermal-limit correlation within the system is weak as expected.
    \item \textbf{Conductance ($\Re(Y^{\mathrm{ser}})$):} Conductance shows a negative correlation ($r \approx -0.27$). In this dataset, branches with larger effective resistive components tend to receive lower branch-importance scores.
\end{enumerate}

To separate the individual contributions of each branch property, we also fit a multivariate regression with conductance, susceptance, thermal limit, and branch type as joint predictors of branch-importance score. Thermal limit has the largest standardized coefficient ($\beta_{\mathrm{std}} \approx 0.36$), followed by susceptance ($\beta_{\mathrm{std}} \approx 0.15$), while conductance has a near-zero independent effect. This pattern is clearer in lines, where resistance and reactance are less correlated than in transformers.

\subsection{Branch Importance Distribution by System}
Fig.~\ref{fig:dist} compares branch-importance distributions across systems. Larger systems show stronger concentration near low branch-importance scores. For \code{case1354pegase}, the median importance is 0.147, and 46.24\% of branches have scores at or below 0.1. In \code{case14} and \code{case30}, the medians are higher (0.297 and 0.298), and the fractions of branches with scores at or below 0.1 are smaller (11.50\% and 15.34\%). For high branch-importance scores, \code{case1354pegase} still has a non-negligible subset: 15.87\% of branches are at or above 0.9, compared with 2.55\% in \code{case14} and 3.29\% in \code{case30}. Intermediate systems such as \code{case118} and \code{case300} also show moderate low-score concentration, with 26.56\% and 24.62\% of branches, respectively, at or below 0.1. Overall, this points to sparser branch weighting in larger networks.

\code{case5} is an outlier among the smaller systems, with a split distribution across branches (30.34\% at or below 0.1 and 22.76\% at or above 0.9). In this system, two of six branches have series susceptance roughly three to five times the median, while the remaining four are tightly clustered near it. With only six branches, each one represents about 17\% of the distribution, so this impedance contrast produces a pronounced split. In comparison, \code{case4gs} has similarly few branches but nearly uniform susceptance, and \code{case14} has comparable impedance spread but 20 branches, so neither shows the same bimodality.

\subsection{Bus Feature Sensitivity and Physics Alignment}
\label{subsec:bus_sensitivity}

\begin{figure*}[t]
    \centering
    \subfloat[Voltage Magnitude ($V_m$) Sensitivity]{\includegraphics[width=0.48\linewidth]{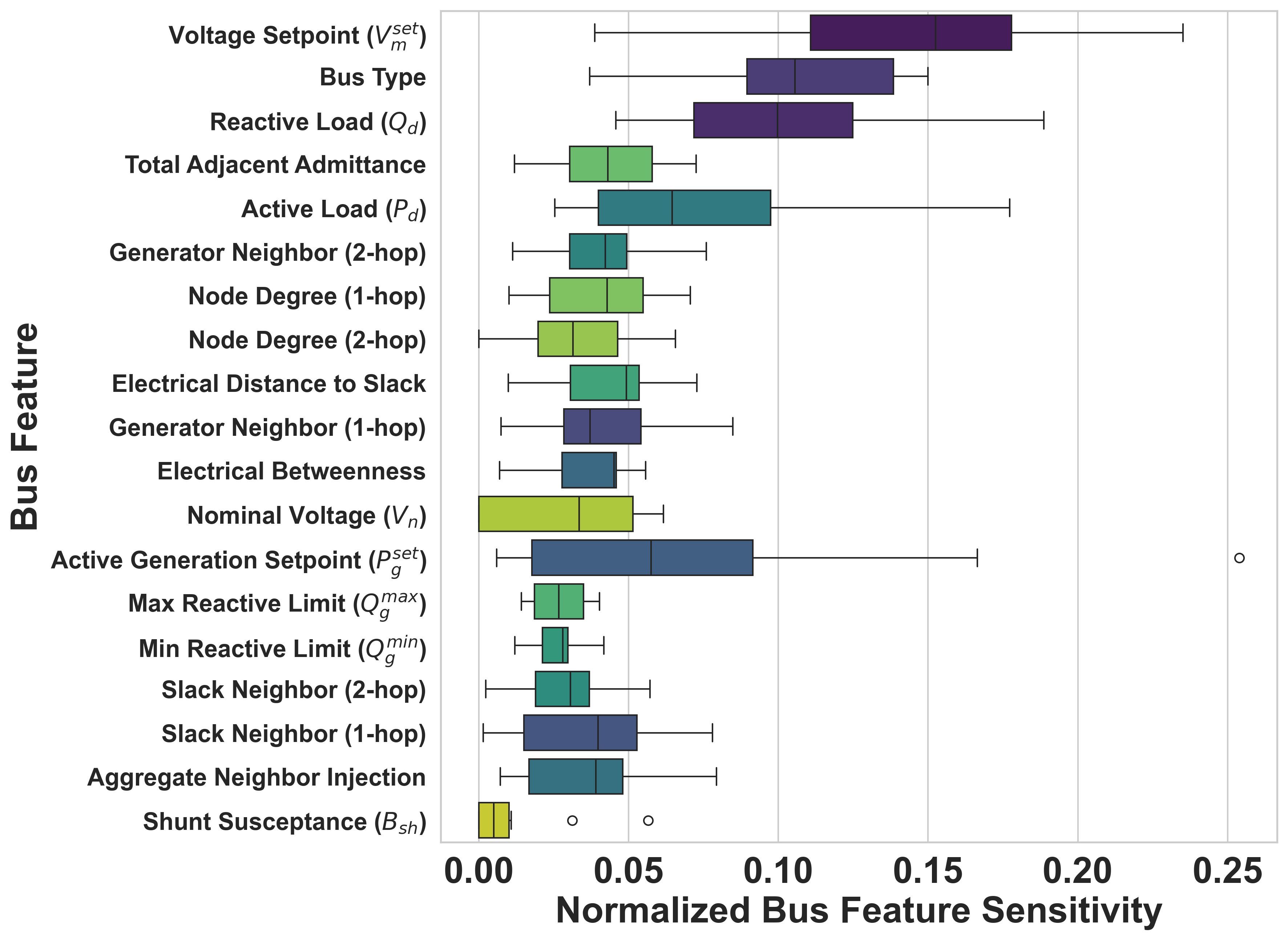}\label{fig:sens_vm}}
    \hfill
    \subfloat[Voltage Angle ($\delta$) Sensitivity]{\includegraphics[width=0.48\linewidth]{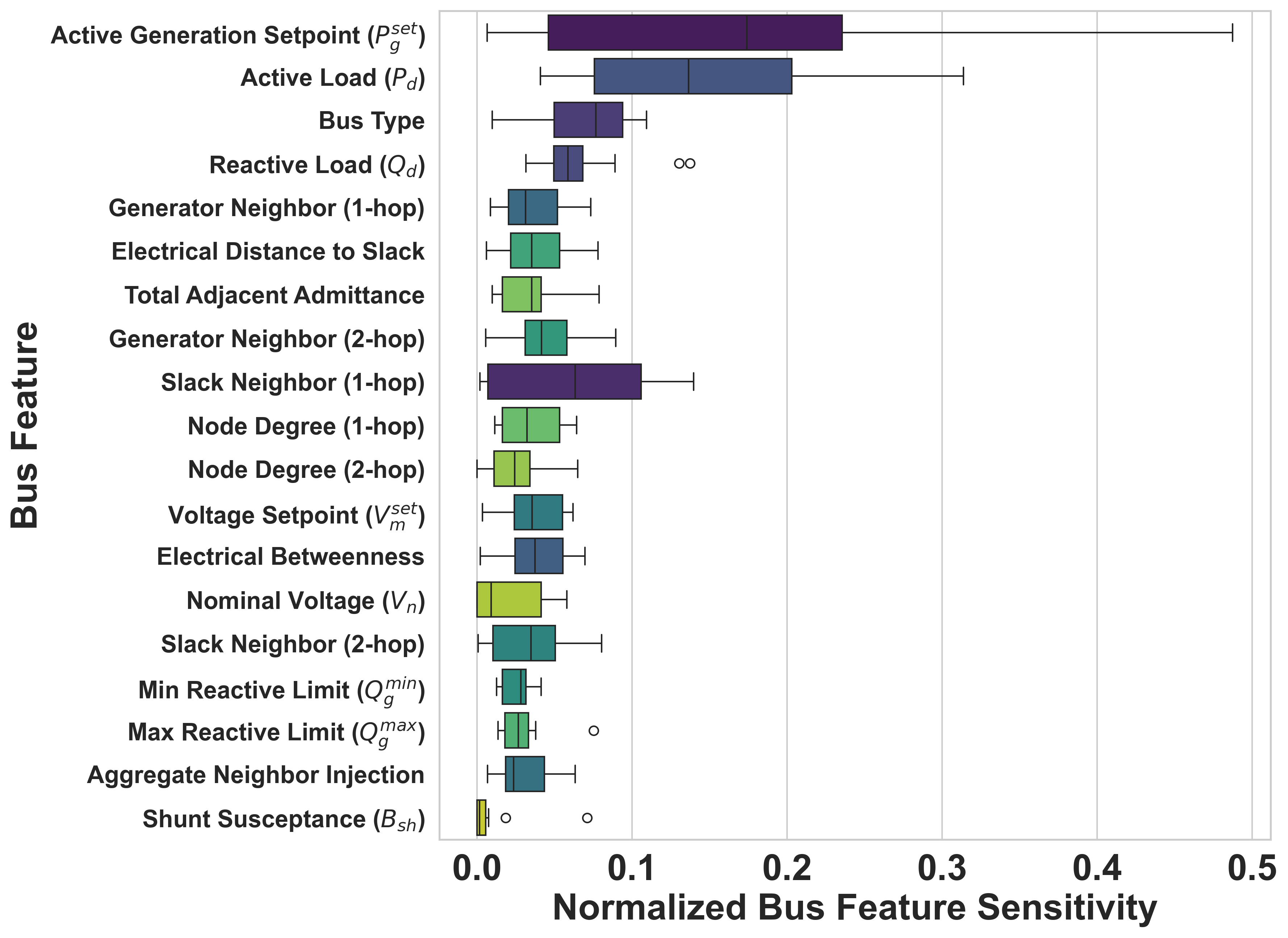}\label{fig:sens_va}}
    \\
    \subfloat[Active Power ($P_g$) Sensitivity]{\includegraphics[width=0.48\linewidth]{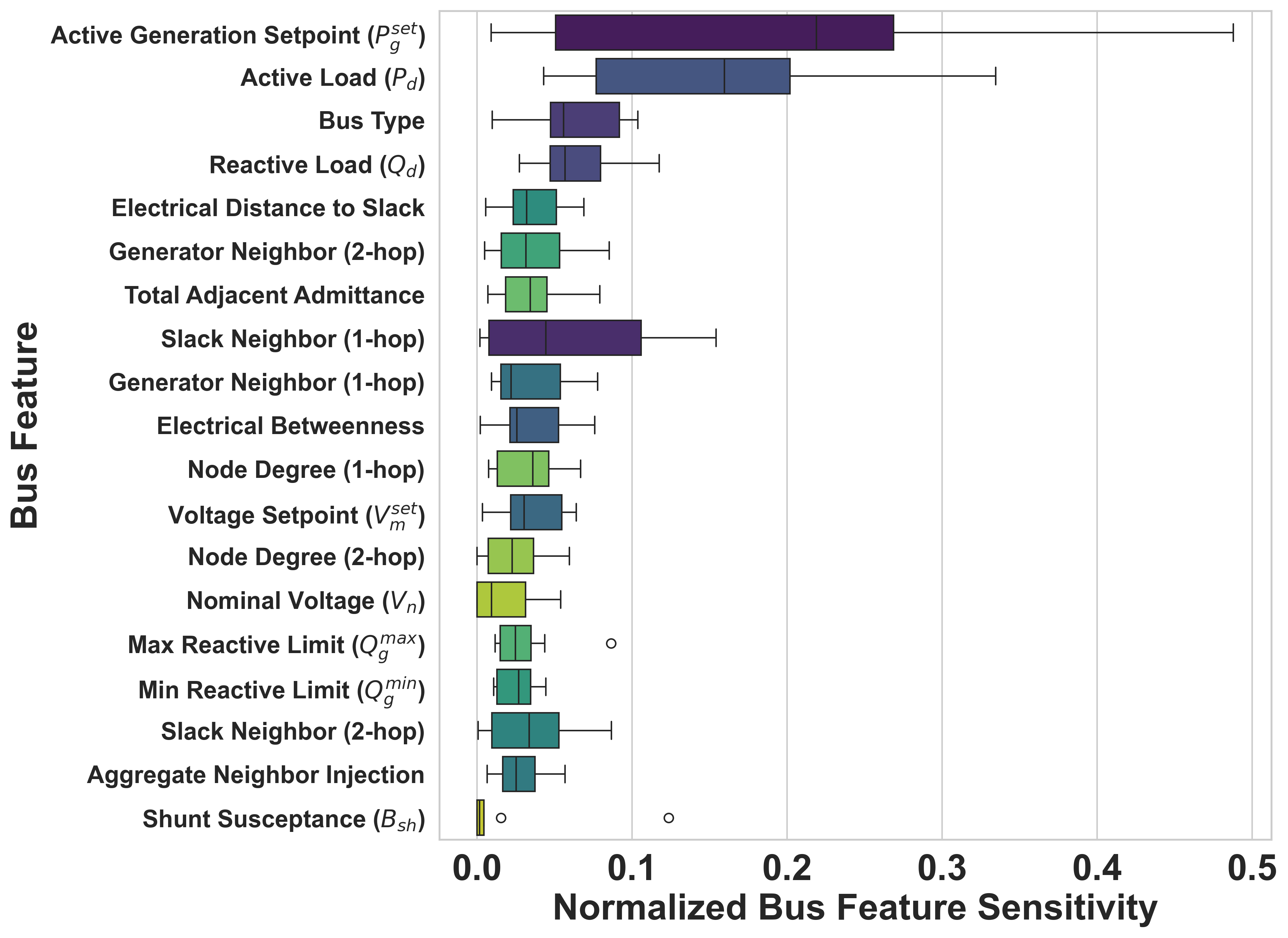}\label{fig:sens_pg}}
    \hfill
    \subfloat[Reactive Power ($Q_g$) Sensitivity]{\includegraphics[width=0.48\linewidth]{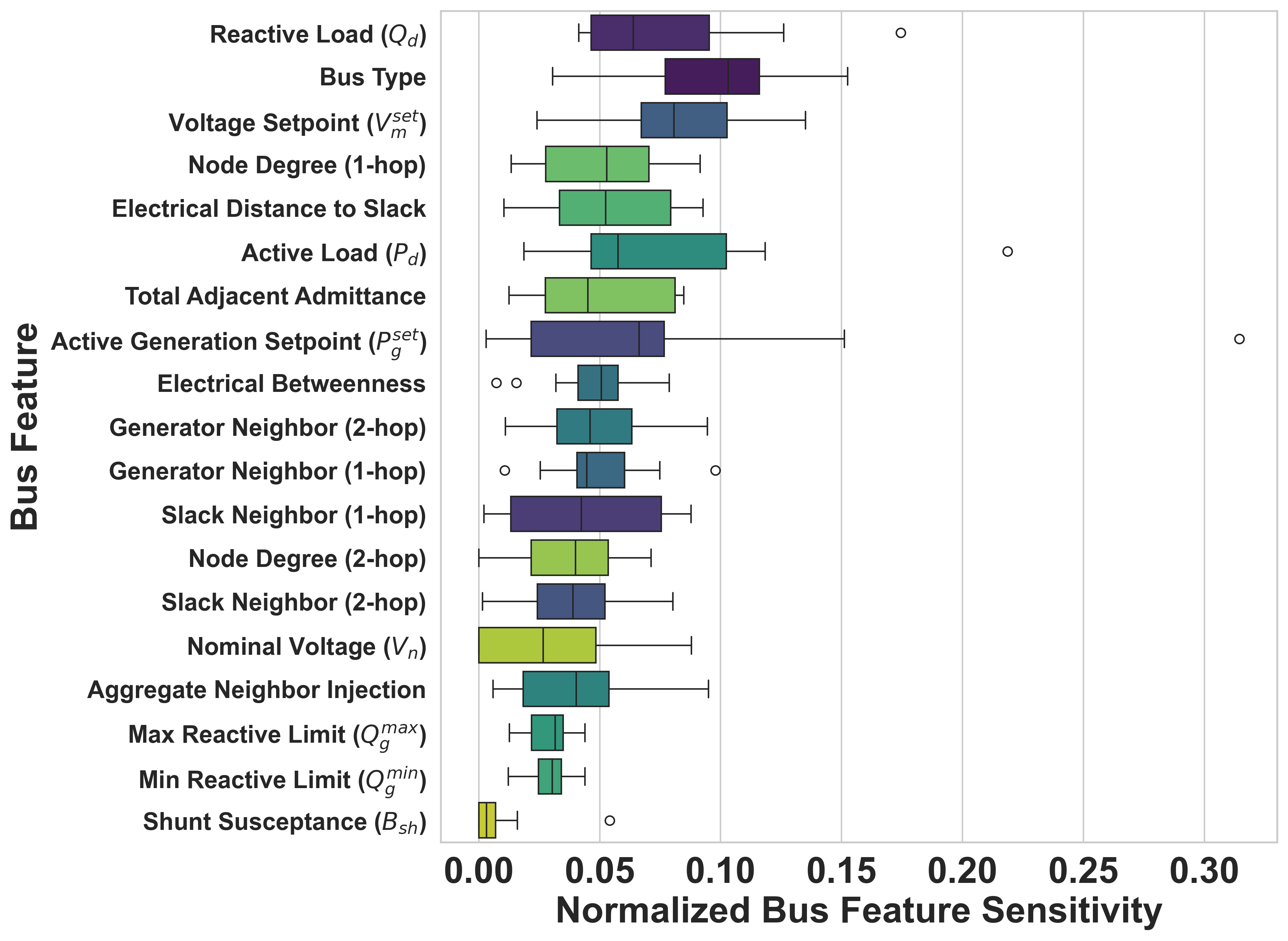}\label{fig:sens_qg}}
    \caption{Global Stability of Bus Feature Sensitivity. Boxplots show the distribution of normalized bus feature sensitivity scores across 13 systems for (a) $V_m$, (b) $\delta$, (c) $P_g$, and (d) $Q_g$.}
    \label{fig:node_sensitivity}
\end{figure*}

Having analyzed edge-level importance, we now examine node-level feature sensitivity. We use integrated gradients (IG)~\cite{sundararajan2017axiomatic} to quantify how input bus features influence each predicted state variable. Using the dataset mean as the baseline, we define the feature-importance score for feature $\phi$ as
\begin{equation}
    \mathcal{I}_{\phi} = \frac{1}{|\mathcal{S}|} \sum_{s \in \mathcal{S}} \left[ \frac{1}{N_s} \sum_{n=1}^{N_s} \left| \mathrm{IG}_{\phi}\!\left(\mathbf{x}_{n}^{(s)}\right) \right| \right],
\end{equation}
where $N_s$ is the total number of node samples from test scenarios in system $s$. We then normalize these scores as $\bar{\mathcal{I}}_{\phi} = \mathcal{I}_{\phi} / \sum_{\phi'} \mathcal{I}_{\phi'}$, so the relative feature importances sum to unity.

Fig.~\ref{fig:node_sensitivity} shows these normalized sensitivities across all 13 systems. At the median level, the top contributors are $V_m^{set}$ for $V_m$ ($\approx 0.15$), $P_g^{set}$ for $\delta$ ($\approx 0.17$), $P_g^{set}$ for $P_g$ ($\approx 0.22$), and Bus Type for $Q_g$ ($\approx 0.10$).

These results show three patterns expected from power system physics:

\subsubsection{$P$-$\delta$ Coupling}
For predicting Voltage Angle ($\delta$) and Active Power Generation ($P_g$) (Figs.~\ref{fig:sens_va} and~\ref{fig:sens_pg}), the primary drivers are Active Generation Setpoint ($P_g^{set}$) and Active Load ($P_d$). For $\delta$, their median normalized sensitivities are $\approx 0.17$ and $\approx 0.14$. For $P_g$, they are $\approx 0.22$ and $\approx 0.16$. This ranking reflects the $P$-$\delta$ coupling in AC power flow, where real-power balance strongly influences voltage angles~\cite{wood2013power}.

\subsubsection{$Q$-$V$ Coupling}
For Voltage Magnitude ($V_m$) prediction (Fig.~\ref{fig:sens_vm}), the Voltage Setpoint ($V_m^{set}$) is the dominant driver (median sensitivity $\approx 0.15$), followed by Bus Type and Reactive Load ($Q_d$). The observed order is expected from the $Q$-$V$ relationship: voltage magnitudes at PV buses are directly controlled by their setpoints, while at PQ buses they respond to local reactive injections~\cite{kundur1994stability}. The prominence of Bus Type suggests that the model distinguishes between bus categories when inferring voltage profiles.

For Reactive Power Generation ($Q_g$) prediction (Fig.~\ref{fig:sens_qg}), Bus Type ($\approx 0.10$) and Voltage Setpoint ($V_m^{set}$, $\approx 0.08$) are the strongest median contributors, followed by Active Generation Setpoint ($P_g^{set}$, $\approx 0.07$). Since $Q_g$ is predicted only at PV and Slack buses (Table~\ref{tab:bus-masks}), this result matches bus-type-specific reactive-power constraints.

\subsubsection{Cross-System Feature Sensitivity}
The stability of these rankings across systems ranging from 14 to 1354 buses supports cross-system consistency. In particular, Bus Type remains stable across targets, with compact IQR values ($\approx 0.05$ for $V_m$, $\approx 0.04$ for $\delta$, and $\approx 0.04$ for $Q_g$), reflecting consistent use of PQ, PV, and Slack bus-type structure across systems.

Together, the branch-importance maps and feature attributions provide a practical diagnostic view of which branches and bus features drive model outputs under different operating conditions.

\section{Conclusion}
\label{sec:conclusion}

This work presented PowerModelsGAT-AI, a unified framework that solves AC power flow across the 13 systems in the trained set using physics-informed graph attention. The key contributions include: (i) system-specific baselines across 14 power systems (4 to 6,470 buses) and a unified multi-system model trained on 13 of these; (ii) a mask-aware formulation that predicts bus voltages and generator injections for all bus types; (iii) a physics-informed loss incorporating power mismatch constraints; and (iv) a continual-learning strategy (EWC+Replay) that mitigates catastrophic forgetting during adaptation to new systems.

Experimental results show that PMGAT-AI achieves strong predictive performance, with an average voltage magnitude NMAE of $0.89\%$ across the unified 13-system benchmark set under the $N\!-\!2$ evaluation regime. PMGAT-AI maintains robust angle prediction, with $R^2 > 0.99$ on the largest transmission systems tested. Analysis of learned attention weights reveals physically meaningful patterns, including stronger concentration of branch importance in larger systems (Sec.~\ref{sec:interpretability}).

Potential directions for future research include extending the contingency dataset and training to higher-order outages (up to $N\!-\!4$), incorporating system security metrics such as the overall performance index (OPI) for contingency ranking and analysis, and adapting the framework to solve the OPF problem. Another direction is to scale unified training to larger systems using gradient checkpointing or distributed training.

\FloatBarrier
\clearpage
\appendices
\raggedbottom

\section{Feature Definitions}
\label{app:features}

This appendix details the node and edge feature sets used in PMGAT-AI. Complete node and edge feature definitions are provided in Table~\ref{tab:node-features} and Table~\ref{tab:edge-features}, respectively.

\begin{table*}[p]
    \centering
    \caption{Node Feature Definitions ($\mathbf{x}_i \in \mathbb{R}^{23}$ + one-hot bus type). Complete specification of bus-level input features used by PowerModelsGAT-AI. See Sec.~\ref{sec:node-edge-features} for details.}
    \label{tab:node-features}
    \renewcommand{\arraystretch}{1.15}
    \footnotesize
    \begin{tabular*}{0.95\textwidth}{@{\extracolsep{\fill}} l c l c @{}}
        \toprule
        \textbf{Feature}             & \textbf{Symbol}    & \textbf{Description}                            & \textbf{Unit} \\
        \midrule
        \multicolumn{4}{@{}l}{\textit{Electrical Properties}}                                                               \\
        \midrule
        Nominal Voltage              & $V_n$              & Rated voltage level of the bus                  & kV            \\
        Active Load                  & $P_d$              & Active power demand at the bus                  & \si{\pu}      \\
        Reactive Load                & $Q_d$              & Reactive power demand at the bus                & \si{\pu}      \\
        Active Generation Setpoint   & $P_g^{\text{set}}$ & Scheduled active power generation               & \si{\pu}      \\
        Voltage Setpoint             & $V_m^{\text{set}}$ & Voltage magnitude setpoint (PV and Slack buses) & \si{\pu}      \\
        Shunt Conductance            & $G_{\text{sh}}$    & Shunt conductance at the bus                    & \si{\pu}      \\
        Shunt Susceptance            & $B_{\text{sh}}$    & Shunt susceptance at the bus                    & \si{\pu}      \\
        \midrule
        \multicolumn{4}{@{}l}{\textit{Physical Limits}}                                                                     \\
        \midrule
        Min Reactive Limit           & $Q_g^{\min}$       & Minimum reactive power generation capacity      & \si{\pu}      \\
        Max Reactive Limit           & $Q_g^{\max}$       & Maximum reactive power generation capacity      & \si{\pu}      \\
        Min Voltage Limit            & $V_m^{\min}$       & Minimum allowable voltage magnitude             & \si{\pu}      \\
        Max Voltage Limit            & $V_m^{\max}$       & Maximum allowable voltage magnitude             & \si{\pu}      \\
        \midrule
        \multicolumn{4}{@{}l}{\textit{System Parameters}}                                                                   \\
        \midrule
        Base Power                   & $S_{\text{base}}$  & System base power for per-unit conversion       & MVA           \\
        Frequency                    & $f$                & System nominal frequency                        & Hz            \\
        \midrule
        \multicolumn{4}{@{}l}{\textit{Topological Features}}                                                                \\
        \midrule
        Node Degree (1-hop)          & ---                & Number of directly connected buses              & count         \\
        Node Degree (2-hop)          & ---                & Number of buses within two hops                 & count         \\
        Electrical Distance to Slack & ---                & Shortest impedance-weighted path length to a slack bus & a.u.          \\
        Electrical Betweenness       & ---                & Betweenness centrality (impedance-weighted)     & normalized    \\
        Aggregate Neighbor Net Injection & ---                & Sum of neighboring net injections               & \si{\pu}      \\
        Total Adjacent Admittance    & ---                & Sum of inverse impedance-based edge weights over incident branches & \si{\pu}      \\
        \midrule
        \multicolumn{4}{@{}l}{\textit{Neighbor Indicators (Binary)}}                                                        \\
        \midrule
        Generator Neighbor (1-hop)   & ---                & Any directly connected bus has a generator      & \{0,1\}       \\
        Generator Neighbor (2-hop)   & ---                & Any 2-hop neighbor has a generator              & \{0,1\}       \\
        Slack Neighbor (1-hop)       & ---                & Any directly connected bus is a slack bus       & \{0,1\}       \\
        Slack Neighbor (2-hop)       & ---                & Any 2-hop neighbor is a slack bus               & \{0,1\}       \\
        \midrule
        \multicolumn{4}{@{}l}{\textit{Categorical}}                                                                         \\
        \midrule
        Bus Type                     & ---                & Power flow bus classification (one-hot encoded) & PQ, PV, Slack   \\
        \bottomrule
    \end{tabular*}
\end{table*}

\begin{table*}[p]
    \centering
    \caption{Edge Feature Definitions ($\mathbf{e}_{ij} \in \mathbb{R}^{7}$). Branch-level attributes for each directed branch edge, encoding series admittance (implemented as directed off-diagonal branch-admittance edge terms), one-hot branch type, and thermal limit.}
    \label{tab:edge-features}
    \renewcommand{\arraystretch}{1.15}
    \footnotesize
    \begin{tabular*}{0.95\textwidth}{@{\extracolsep{\fill}} l c l c @{}}
        \toprule
        \textbf{Feature} & \textbf{Symbol} & \textbf{Description}                & \textbf{Unit} \\
        \midrule
        \multicolumn{4}{@{}l}{\textit{Series Admittance (Edge Terms)}}                                        \\
        \midrule
        Conductance      & $G$             & Real part of the series-admittance edge term      & \si{\pu}      \\
        Susceptance      & $B$             & Imaginary part of the series-admittance edge term & \si{\pu}      \\
        \midrule
        \multicolumn{4}{@{}l}{\textit{Branch Type (One-Hot)}}                                    \\
        \midrule
        Line             & ---             & Transmission line                   & \{0,1\}       \\
        Transformer      & ---             & Power transformer                   & \{0,1\}       \\
        Impedance        & ---             & Series impedance element            & \{0,1\}       \\
        Switch           & ---             & Switching device                    & \{0,1\}       \\
        \midrule
        \multicolumn{4}{@{}l}{\textit{Thermal Limit}}                                           \\
        \midrule
        Thermal Limit    & $I^{\text{max}}$      & Per-unit thermal-limit proxy computed as branch rating (MVA) normalized by system base power & \si{\pu}      \\
        \bottomrule
    \end{tabular*}
\end{table*}

\FloatBarrier
\clearpage
\section{Dataset Contingency Statistics}
\label{app:contingency-stats}

This appendix reports contingency statistics of the dataset. Table~\ref{tab:bus-stats} lists the static bus-type composition per system, and Tables~\ref{tab:contingency-specialized} and~\ref{tab:contingency-unified} summarize the training and test contingency distributions. The $N\!-\!k$ label gives the number of removed branches per scenario: $N\!-\!0$ (normal), $N\!-\!1$ (single-branch outage), and $N\!-\!2$ (double-branch outage). For \code{case4gs} and \code{case9}, $N\!-\!2$ samples are absent (see Sec.~\ref{sec:setup}).

\begin{table*}[p]
    \centering
    \caption{Bus Type Distribution by System. Breakdown of bus counts per system, classified by bus type (PQ, PV, and Slack).}
    \label{tab:bus-stats}
    \renewcommand{\arraystretch}{1.0}
    \setlength{\tabcolsep}{0pt}
    \footnotesize
    \begin{tabular*}{0.95\textwidth}{@{\extracolsep{\fill}} l c c c c @{}}
        \toprule
        \textbf{System}   & \textbf{Total} & \textbf{PQ (Load)} & \textbf{PV (Gen)} & \textbf{Slack} \\
        \midrule
        case4gs           & 4              & 2                  & 1                 & 1              \\
        case5             & 5              & 1                  & 3                 & 1              \\
        case6ww           & 6              & 3                  & 2                 & 1              \\
        case9             & 9              & 6                  & 2                 & 1              \\
        case14            & 14             & 9                  & 4                 & 1              \\
        case30            & 30             & 24                 & 5                 & 1              \\
        case\_ieee30      & 30             & 24                 & 5                 & 1              \\
        case39            & 39             & 29                 & 9                 & 1              \\
        case57            & 57             & 50                 & 6                 & 1              \\
        case118           & 118            & 64                 & 53                & 1              \\
        case\_illinois200 & 200            & 162                & 37                & 1              \\
        case300           & 300            & 231                & 68                & 1              \\
        case1354pegase    & 1354           & 1094               & 259               & 1              \\
        case6470rte       & 6470           & 6017               & 452               & 1              \\
        \bottomrule
    \end{tabular*}
\end{table*}

\begin{table*}[p]
    \centering
    \caption{Contingency Distribution for System-Specific Models ($N\!-\!2$ Training and Test Sets).}
    \label{tab:contingency-specialized}
    \renewcommand{\arraystretch}{1.1}
    \setlength{\tabcolsep}{0pt}
    \footnotesize
    \begin{tabular*}{0.95\textwidth}{@{\extracolsep{\fill}} l c c c c c c @{}}
        \toprule
        & \multicolumn{2}{c}{\textbf{N-0 (\%)}} & \multicolumn{2}{c}{\textbf{N-1 (\%)}} & \multicolumn{2}{c}{\textbf{N-2 (\%)}} \\
        \cmidrule(lr){2-3} \cmidrule(lr){4-5} \cmidrule(lr){6-7}
        \textbf{System}   & \textbf{Train} & \textbf{Test} & \textbf{Train} & \textbf{Test} & \textbf{Train} & \textbf{Test} \\
        \midrule
        case4gs           & 49.5 & 50.1 & 50.5 & 49.9 & --- & --- \\
        case5             & 36.1 & 33.7 & 35.7 & 34.4 & 28.2 & 31.9 \\
        case6ww           & 35.0 & 32.1 & 33.1 & 33.2 & 31.9 & 34.7 \\
        case9             & 60.5 & 60.0 & 39.5 & 40.0 & --- & --- \\
        case14            & 36.3 & 35.4 & 33.7 & 34.4 & 30.0 & 30.2 \\
        case30            & 36.3 & 38.5 & 34.1 & 33.2 & 29.6 & 28.3 \\
        case\_ieee30      & 37.6 & 37.4 & 32.9 & 34.0 & 29.5 & 28.6 \\
        case39            & 44.8 & 42.7 & 33.3 & 32.5 & 21.9 & 24.8 \\
        case57            & 36.3 & 35.3 & 33.0 & 33.9 & 30.7 & 30.8 \\
        case118           & 34.2 & 34.9 & 34.1 & 32.7 & 31.7 & 32.4 \\
        case\_illinois200 & 46.2 & 45.5 & 31.7 & 31.6 & 22.1 & 22.9 \\
        case300           & 43.6 & 39.9 & 32.0 & 36.0 & 24.4 & 24.1 \\
        case1354pegase    & 44.0 & 45.7 & 32.3 & 30.6 & 23.7 & 23.7 \\
        case6470rte       & 44.6 & 43.3 & 33.0 & 31.0 & 22.4 & 25.7 \\
        \bottomrule
    \end{tabular*}
\end{table*}

\begin{table*}[p]
    \centering
    \caption{Contingency Distribution for Unified Models ($N\!-\!2$ Training and Test Sets). Each system contributes 2,000 scenarios. Comparisons show the stability of the contingency distribution across training and test sets.}
    \label{tab:contingency-unified}
    \renewcommand{\arraystretch}{1.1}
    \setlength{\tabcolsep}{0pt}
    \footnotesize
    \begin{tabular*}{0.95\textwidth}{@{\extracolsep{\fill}} l c c c c c c @{}}
        \toprule
        & \multicolumn{2}{c}{\textbf{N-0 (\%)}} & \multicolumn{2}{c}{\textbf{N-1 (\%)}} & \multicolumn{2}{c}{\textbf{N-2 (\%)}}                                                  \\
        \cmidrule(lr){2-3} \cmidrule(lr){4-5} \cmidrule(lr){6-7}
        \textbf{Model}    & \textbf{Train}                        & \textbf{Test}                         & \textbf{Train}                        & \textbf{Test} & \textbf{Train} & \textbf{Test} \\
        \midrule
        Unified (13 Systems) & 40.3                                  & 41.8                                  & 36.2                                  & 35.1          & 23.5           & 23.1          \\
        Base (12 Systems)    & 39.5                                  & 41.5                                  & 37.3                                  & 35.6          & 23.2           & 22.8          \\
        \bottomrule
    \end{tabular*}
\end{table*}

\FloatBarrier
\clearpage
\section{System-Specific Baseline Results}
\label{app:baselines}

This appendix provides detailed baseline performance metrics for each system under normal ($N\!-\!0$) and contingency ($N\!-\!2$) conditions.

\begin{table*}[p]
    \centering
    \caption{System-Specific Baseline Performance ($N\!-\!0$). Per-system metrics for system-specific models described in Sec.~\ref{subsec:n0-results}. Each model is trained and evaluated on its respective system under normal operating conditions (no element outages).}
    \label{tab:single-topo-n0}
    \renewcommand{\arraystretch}{1.1}
    \setlength{\tabcolsep}{0pt}
    \footnotesize
    \begin{tabular*}{0.95\textwidth}{@{\extracolsep{\fill}}
            l
            l
            S[round-mode=figures, round-precision=3, exponent-mode=scientific]
            S[round-mode=places, round-precision=4]
            S[round-mode=places, round-precision=4]
            S[round-mode=places, round-precision=4]
            S[round-mode=places, round-precision=4]
            @{}}
        \toprule
        \textbf{System} & \textbf{Target}         & {\textbf{MSE}} & {\textbf{RMSE}} & {\textbf{MAE}} & {\textbf{R\textsuperscript{2}}} & {\textbf{NMAE\%}} \\
        \midrule
        \multirow{4}{*}{\textbf{case4gs}}
        & $V_m$ [\si{\pu}]        & 1.134e-07      & 3.368e-04       & 2.672e-04      & 0.9996           & 0.2652            \\
        & $\delta$ [\si{\degree}] & 4.121e-04      & 0.02030         & 0.01481        & 0.9999           & 0.1274            \\
        & $P_g$ [\si{\pu}]        & 6.434e-04      & 0.02536         & 0.01849        & 0.9996           & 0.2794            \\
        & $Q_g$ [\si{\pu}]        & 5.654e-04      & 0.02378         & 0.01832        & 0.9995           & 0.3571            \\
        \midrule
        \multirow{4}{*}{\textbf{case5}}
        & $V_m$ [\si{\pu}]        & 1.290e-07      & 3.592e-04       & 3.017e-04      & 0.9996           & 0.3619            \\
        & $\delta$ [\si{\degree}] & 0.001159       & 0.03405         & 0.02628        & 0.9998           & 0.2556            \\
        & $P_g$ [\si{\pu}]        & 0.002345       & 0.04842         & 0.03840        & 0.9994           & 0.3747            \\
        & $Q_g$ [\si{\pu}]        & 0.01551        & 0.1245          & 0.09150        & 0.9993           & 0.2829            \\
        \midrule
        \multirow{4}{*}{\textbf{case6ww}}
        & $V_m$ [\si{\pu}]        & 2.935e-07      & 5.418e-04       & 4.285e-04      & 0.9995           & 0.2722            \\
        & $\delta$ [\si{\degree}] & 9.686e-04      & 0.03112         & 0.02383        & 0.9999           & 0.1208            \\
        & $P_g$ [\si{\pu}]        & 3.147e-05      & 0.005610        & 0.004179       & 0.9999           & 0.1475            \\
        & $Q_g$ [\si{\pu}]        & 1.152e-04      & 0.01073         & 0.008237       & 0.9995           & 0.2266            \\
        \midrule
        \multirow{4}{*}{\textbf{case9}}
        & $V_m$ [\si{\pu}]        & 1.395e-06      & 0.001181        & 9.378e-04      & 0.9983           & 0.5361            \\
        & $\delta$ [\si{\degree}] & 0.002577       & 0.05076         & 0.03902        & 0.9999           & 0.0887            \\
        & $P_g$ [\si{\pu}]        & 1.261e-04      & 0.01123         & 0.008149       & 0.9998           & 0.2153            \\
        & $Q_g$ [\si{\pu}]        & 2.511e-04      & 0.01585         & 0.01283        & 0.9943           & 0.9679            \\
        \midrule
        \multirow{4}{*}{\textbf{case14}}
        & $V_m$ [\si{\pu}]        & 1.748e-07      & 4.181e-04       & 3.279e-04      & 0.9996           & 0.2398            \\
        & $\delta$ [\si{\degree}] & 0.002905       & 0.05390         & 0.04136        & 0.9999           & 0.1623            \\
        & $P_g$ [\si{\pu}]        & 5.692e-05      & 0.007545        & 0.005625       & 0.9998           & 0.1823            \\
        & $Q_g$ [\si{\pu}]        & 1.912e-04      & 0.01383         & 0.009716       & 0.9993           & 0.2751            \\
        \midrule
        \multirow{4}{*}{\textbf{case30}}
        & $V_m$ [\si{\pu}]        & 2.735e-07      & 5.230e-04       & 4.047e-04      & 0.9992           & 0.3105            \\
        & $\delta$ [\si{\degree}] & 0.003275       & 0.05723         & 0.04185        & 0.9997           & 0.1779            \\
        & $P_g$ [\si{\pu}]        & 4.123e-05      & 0.006421        & 0.004762       & 0.9998           & 0.2230            \\
        & $Q_g$ [\si{\pu}]        & 2.017e-04      & 0.01420         & 0.009114       & 0.9995           & 0.2045            \\
        \midrule
        \multirow{4}{*}{\textbf{case\_ieee30}}
        & $V_m$ [\si{\pu}]        & 4.821e-07      & 6.944e-04       & 5.270e-04      & 0.9990           & 0.3162            \\
        & $\delta$ [\si{\degree}] & 0.04552        & 0.2134          & 0.1668         & 0.9979           & 0.5688            \\
        & $P_g$ [\si{\pu}]        & 1.861e-04      & 0.01364         & 0.01083        & 0.9996           & 0.3300            \\
        & $Q_g$ [\si{\pu}]        & 4.971e-04      & 0.02230         & 0.01527        & 0.9984           & 0.4112            \\
        \midrule
        \multirow{4}{*}{\textbf{case39}}
        & $V_m$ [\si{\pu}]        & 2.339e-06      & 0.001529        & 0.001074       & 0.9980           & 0.3135            \\
        & $\delta$ [\si{\degree}] & 0.1848         & 0.4299          & 0.2590         & 0.9998           & 0.1456            \\
        & $P_g$ [\si{\pu}]        & 0.01386        & 0.1177          & 0.09067        & 0.9999           & 0.1930            \\
        & $Q_g$ [\si{\pu}]        & 0.01080        & 0.1039          & 0.06830        & 0.9975           & 0.2756            \\
        \midrule
        \multirow{4}{*}{\textbf{case57}}
        & $V_m$ [\si{\pu}]        & 2.629e-06      & 0.001622        & 0.001138       & 0.9996           & 0.2048            \\
        & $\delta$ [\si{\degree}] & 0.03025        & 0.1739          & 0.1273         & 0.9996           & 0.1835            \\
        & $P_g$ [\si{\pu}]        & 0.001378       & 0.03712         & 0.02948        & 0.9999           & 0.1948            \\
        & $Q_g$ [\si{\pu}]        & 0.002122       & 0.04607         & 0.03621        & 0.9982           & 0.5355            \\
        \midrule
        \multirow{4}{*}{\textbf{case118}}
        & $V_m$ [\si{\pu}]        & 5.940e-07      & 7.707e-04       & 5.771e-04      & 0.9984           & 0.3972            \\
        & $\delta$ [\si{\degree}] & 0.1797         & 0.4239          & 0.3186         & 0.9995           & 0.2163            \\
        & $P_g$ [\si{\pu}]        & 0.04669        & 0.2161          & 0.1912         & 0.9995           & 0.4364            \\
        & $Q_g$ [\si{\pu}]        & 0.008408       & 0.09169         & 0.06636        & 0.9976           & 0.2370            \\
        \midrule
        \multirow{4}{*}{\textbf{case\_illinois200}}
        & $V_m$ [\si{\pu}]        & 7.409e-07      & 8.607e-04       & 6.508e-04      & 0.9984           & 0.4159            \\
        & $\delta$ [\si{\degree}] & 0.09201        & 0.3033          & 0.2262         & 0.9990           & 0.3216            \\
        & $P_g$ [\si{\pu}]        & 0.003798       & 0.06163         & 0.05383        & 0.9999           & 0.2185            \\
        & $Q_g$ [\si{\pu}]        & 2.261e-04      & 0.01504         & 0.009167       & 0.9985           & 0.1211            \\
        \midrule
        \multirow{4}{*}{\textbf{case300}}
        & $V_m$ [\si{\pu}]        & 6.682e-05      & 0.008175        & 0.003978       & 0.9747           & 0.6270            \\
        & $\delta$ [\si{\degree}] & 4.029          & 2.007           & 1.478          & 0.9978           & 0.4153            \\
        & $P_g$ [\si{\pu}]        & 0.04394        & 0.2096          & 0.1579         & 0.9991           & 0.5945            \\
        & $Q_g$ [\si{\pu}]        & 0.01592        & 0.1262          & 0.08529        & 0.9979           & 0.1818            \\
        \midrule
        \multirow{4}{*}{\textbf{case1354pegase}}
        & $V_m$ [\si{\pu}]        & 1.804e-05      & 0.004247        & 0.002802       & 0.9725           & 0.8564            \\
        & $\delta$ [\si{\degree}] & 6.861          & 2.619           & 1.955          & 0.9972           & 0.6158            \\
        & $P_g$ [\si{\pu}]        & 8.627          & 2.937           & 2.570          & 0.9994           & 0.5451            \\
        & $Q_g$ [\si{\pu}]        & 3.984          & 1.996           & 1.098          & 0.9966           & 0.1077            \\
        \midrule
        \multirow{4}{*}{\textbf{case6470rte}}
        & $V_m$ [\si{\pu}]        & 7.382e-05      & 0.008592        & 0.006504       & 0.9441           & 0.9846            \\
        & $\delta$ [\si{\degree}] & 53.26          & 7.298           & 4.968          & 0.9892           & 1.3800            \\
        & $P_g$ [\si{\pu}]        & 6.071          & 2.464           & 2.002          & 0.9981           & 0.9568            \\
        & $Q_g$ [\si{\pu}]        & 0.1259         & 0.3548          & 0.1721         & 0.9909           & 0.1483            \\
        \bottomrule
    \end{tabular*}
\end{table*}

\begin{table*}[p]
    \centering
    \caption{System-Specific Contingency Performance ($N\!-\!2$). Per-system metrics for system-specific models described in Sec.~\ref{subsec:n2-results}. Each model is trained and evaluated on scenarios with up to two simultaneous branch outages, testing robustness under severe ($N\!-\!2$) contingencies.}
    \label{tab:single-topo-n2}
    \renewcommand{\arraystretch}{1.05}
    \setlength{\tabcolsep}{0pt}
    \footnotesize
    \begin{tabular*}{0.95\textwidth}{@{\extracolsep{\fill}}
            l
            l
            S[round-mode=figures, round-precision=3, exponent-mode=scientific]
            S[round-mode=places, round-precision=4]
            S[round-mode=places, round-precision=4]
            S[round-mode=places, round-precision=4]
            S[round-mode=places, round-precision=4]
            @{}}
        \toprule
        \textbf{System} & \textbf{Target}         & {\textbf{MSE}} & {\textbf{RMSE}} & {\textbf{MAE}} & {\textbf{R\textsuperscript{2}}} & {\textbf{NMAE\%}} \\
        \midrule
        \multirow{4}{*}{\textbf{case4gs}}
        & $V_m$ [\si{\pu}]        & 2.597e-07      & 5.096e-04       & 3.502e-04      & 0.9997           & 0.1353            \\
        & $\delta$ [\si{\degree}] & 0.002268       & 0.04762         & 0.03280        & 0.9999           & 0.0763            \\
        & $P_g$ [\si{\pu}]        & 3.522e-04      & 0.01877         & 0.01517        & 0.9998           & 0.2552            \\
        & $Q_g$ [\si{\pu}]        & 3.339e-04      & 0.01827         & 0.01441        & 0.9997           & 0.2031            \\
        \midrule
        \multirow{4}{*}{\textbf{case5}}
        & $V_m$ [\si{\pu}]        & 2.116e-07      & 4.600e-04       & 3.145e-04      & 0.9996           & 0.2614            \\
        & $\delta$ [\si{\degree}] & 0.03161        & 0.1778          & 0.09841        & 0.9987           & 0.2410            \\
        & $P_g$ [\si{\pu}]        & 0.002953       & 0.05434         & 0.04474        & 0.9993           & 0.4343            \\
        & $Q_g$ [\si{\pu}]        & 0.01228        & 0.1108          & 0.08139        & 0.9993           & 0.2711            \\
        \midrule
        \multirow{4}{*}{\textbf{case6ww}}
        & $V_m$ [\si{\pu}]        & 4.495e-06      & 0.002120        & 7.581e-04      & 0.9977           & 0.1698            \\
        & $\delta$ [\si{\degree}] & 0.1000         & 0.3162          & 0.08097        & 0.9965           & 0.1325            \\
        & $P_g$ [\si{\pu}]        & 2.264e-04      & 0.01505         & 0.009437       & 0.9993           & 0.2903            \\
        & $Q_g$ [\si{\pu}]        & 3.512e-04      & 0.01874         & 0.01103        & 0.9990           & 0.2135            \\
        \midrule
        \multirow{4}{*}{\textbf{case9}}
        & $V_m$ [\si{\pu}]        & 2.578e-06      & 0.001606        & 0.001075       & 0.9982           & 0.2795            \\
        & $\delta$ [\si{\degree}] & 0.01784        & 0.1336          & 0.07721        & 0.9998           & 0.0592            \\
        & $P_g$ [\si{\pu}]        & 1.150e-04      & 0.01073         & 0.008847       & 0.9998           & 0.2245            \\
        & $Q_g$ [\si{\pu}]        & 3.321e-04      & 0.01822         & 0.01387        & 0.9955           & 0.6693            \\
        \midrule
        \multirow{4}{*}{\textbf{case14}}
        & $V_m$ [\si{\pu}]        & 2.792e-06      & 0.001671        & 7.185e-04      & 0.9955           & 0.1475            \\
        & $\delta$ [\si{\degree}] & 0.1173         & 0.3424          & 0.1739         & 0.9977           & 0.2519            \\
        & $P_g$ [\si{\pu}]        & 3.380e-04      & 0.01839         & 0.01307        & 0.9991           & 0.4077            \\
        & $Q_g$ [\si{\pu}]        & 2.854e-04      & 0.01689         & 0.01144        & 0.9989           & 0.2854            \\
        \midrule
        \multirow{4}{*}{\textbf{case30}}
        & $V_m$ [\si{\pu}]        & 7.611e-07      & 8.724e-04       & 6.365e-04      & 0.9982           & 0.2340            \\
        & $\delta$ [\si{\degree}] & 0.01802        & 0.1342          & 0.07863        & 0.9988           & 0.2108            \\
        & $P_g$ [\si{\pu}]        & 6.472e-05      & 0.008045        & 0.005644       & 0.9997           & 0.2447            \\
        & $Q_g$ [\si{\pu}]        & 3.211e-04      & 0.01792         & 0.01171        & 0.9992           & 0.2546            \\
        \midrule
        \multirow{4}{*}{\textbf{case\_ieee30}}
        & $V_m$ [\si{\pu}]        & 2.870e-06      & 0.001694        & 0.001085       & 0.9960           & 0.3040            \\
        & $\delta$ [\si{\degree}] & 0.2736         & 0.5231          & 0.2767         & 0.9933           & 0.3963            \\
        & $P_g$ [\si{\pu}]        & 3.698e-04      & 0.01923         & 0.01228        & 0.9992           & 0.3618            \\
        & $Q_g$ [\si{\pu}]        & 6.803e-04      & 0.02608         & 0.01673        & 0.9978           & 0.4454            \\
        \midrule
        \multirow{4}{*}{\textbf{case39}}
        & $V_m$ [\si{\pu}]        & 7.331e-06      & 0.002708        & 0.001575       & 0.9943           & 0.4375            \\
        & $\delta$ [\si{\degree}] & 3.577          & 1.891           & 0.5171         & 0.9958           & 0.2216            \\
        & $P_g$ [\si{\pu}]        & 0.007932       & 0.08906         & 0.06492        & 0.9999           & 0.1374            \\
        & $Q_g$ [\si{\pu}]        & 0.02504        & 0.1582          & 0.08314        & 0.9942           & 0.3605            \\
        \midrule
        \multirow{4}{*}{\textbf{case57}}
        & $V_m$ [\si{\pu}]        & 1.238e-05      & 0.003518        & 0.001935       & 0.9982           & 0.3096            \\
        & $\delta$ [\si{\degree}] & 0.1076         & 0.3281          & 0.2052         & 0.9988           & 0.1984            \\
        & $P_g$ [\si{\pu}]        & 0.006119       & 0.07822         & 0.05863        & 0.9993           & 0.4110            \\
        & $Q_g$ [\si{\pu}]        & 0.004132       & 0.06428         & 0.05093        & 0.9965           & 0.7165            \\
        \midrule
        \multirow{4}{*}{\textbf{case118}}
        & $V_m$ [\si{\pu}]        & 9.187e-07      & 9.585e-04       & 6.478e-04      & 0.9976           & 0.3425            \\
        & $\delta$ [\si{\degree}] & 0.4869         & 0.6978          & 0.4666         & 0.9989           & 0.2883            \\
        & $P_g$ [\si{\pu}]        & 0.04802        & 0.2191          & 0.2020         & 0.9995           & 0.4449            \\
        & $Q_g$ [\si{\pu}]        & 0.008382       & 0.09155         & 0.06775        & 0.9976           & 0.2423            \\
        \midrule
        \multirow{4}{*}{\textbf{case\_illinois200}}
        & $V_m$ [\si{\pu}]        & 2.812e-06      & 0.001677        & 0.001111       & 0.9945           & 0.4120            \\
        & $\delta$ [\si{\degree}] & 0.2153         & 0.4640          & 0.2957         & 0.9979           & 0.3041            \\
        & $P_g$ [\si{\pu}]        & 0.002042       & 0.04519         & 0.03299        & 0.9999           & 0.1307            \\
        & $Q_g$ [\si{\pu}]        & 3.278e-04      & 0.01811         & 0.01015        & 0.9979           & 0.1157            \\
        \midrule
        \multirow{4}{*}{\textbf{case300}}
        & $V_m$ [\si{\pu}]        & 5.132e-05      & 0.007164        & 0.003800       & 0.9811           & 0.5010            \\
        & $\delta$ [\si{\degree}] & 8.368          & 2.893           & 1.952          & 0.9956           & 0.5713            \\
        & $P_g$ [\si{\pu}]        & 0.06041        & 0.2458          & 0.1758         & 0.9988           & 0.6712            \\
        & $Q_g$ [\si{\pu}]        & 0.02648        & 0.1627          & 0.1013         & 0.9965           & 0.2337            \\
        \midrule
        \multirow{4}{*}{\textbf{case1354pegase}}
        & $V_m$ [\si{\pu}]        & 1.345e-05      & 0.003667        & 0.002452       & 0.9792           & 0.5358            \\
        & $\delta$ [\si{\degree}] & 4.615          & 2.148           & 1.600          & 0.9980           & 0.5098            \\
        & $P_g$ [\si{\pu}]        & 2.470          & 1.572           & 1.358          & 0.9998           & 0.2931            \\
        & $Q_g$ [\si{\pu}]        & 5.015          & 2.239           & 1.140          & 0.9957           & 0.1166            \\
        \midrule
        \multirow{4}{*}{\textbf{case6470rte}}
        & $V_m$ [\si{\pu}]        & 9.468e-05      & 0.009730        & 0.007333       & 0.9268           & 1.1110            \\
        & $\delta$ [\si{\degree}] & 36.16          & 6.013           & 4.358          & 0.9918           & 1.2110            \\
        & $P_g$ [\si{\pu}]        & 2.896          & 1.702           & 1.345          & 0.9990           & 0.6357            \\
        & $Q_g$ [\si{\pu}]        & 0.1450         & 0.3807          & 0.2012         & 0.9893           & 0.1641            \\
        \bottomrule
    \end{tabular*}
\end{table*}

\FloatBarrier
\clearpage
\section{Unified Model Per-System Results}
\label{app:unified}

This appendix reports the per-system performance of the unified model, which uses a single set of shared weights to solve the AC power flow across all 13 systems.

\begin{table*}[p]
    \centering
    \caption{Unified PowerModelsGAT-AI Performance Across All 13 Systems ($N\!-\!2$ Contingencies). Per-system breakdown for results discussed in Sec.~\ref{subsec:unified-results}.}
    \label{tab:unified-n2}
    \renewcommand{\arraystretch}{1.1}
    \setlength{\tabcolsep}{0pt}
    \footnotesize
    \begin{tabular*}{0.95\textwidth}{@{\extracolsep{\fill}}
            l
            l
            S[round-mode=figures, round-precision=3, exponent-mode=scientific]
            S[round-mode=places, round-precision=4]
            S[round-mode=places, round-precision=4]
            S[round-mode=places, round-precision=4]
            S[round-mode=places, round-precision=4]
            @{}}
        \toprule
        \textbf{System} & \textbf{Target}         & {\textbf{MSE}} & {\textbf{RMSE}} & {\textbf{MAE}} & {\textbf{R\textsuperscript{2}}} & {\textbf{NMAE\%}} \\
        \midrule
        \multirow{4}{*}{\textbf{case4gs}}
        & $V_m$ [\si{\pu}]        & 2.147e-06      & 0.001465        & 0.001032       & 0.9978           & 0.5636            \\
        & $\delta$ [\si{\degree}] & 0.1484         & 0.3852          & 0.2606         & 0.9938           & 0.6885            \\
        & $P_g$ [\si{\pu}]        & 0.006864       & 0.08285         & 0.05996        & 0.9953           & 1.0250            \\
        & $Q_g$ [\si{\pu}]        & 0.009688       & 0.09843         & 0.07650        & 0.9914           & 1.2720            \\
        \midrule
        \multirow{4}{*}{\textbf{case5}}
        & $V_m$ [\si{\pu}]        & 2.283e-06      & 0.001511        & 0.001147       & 0.9946           & 1.1010            \\
        & $\delta$ [\si{\degree}] & 0.4478         & 0.6692          & 0.4484         & 0.9832           & 1.0560            \\
        & $P_g$ [\si{\pu}]        & 0.03883        & 0.1971          & 0.1236         & 0.9919           & 1.2330            \\
        & $Q_g$ [\si{\pu}]        & 0.1450         & 0.3807          & 0.2781         & 0.9911           & 0.9824            \\
        \midrule
        \multirow{4}{*}{\textbf{case6ww}}
        & $V_m$ [\si{\pu}]        & 5.255e-06      & 0.002292        & 0.001641       & 0.9964           & 0.5615            \\
        & $\delta$ [\si{\degree}] & 0.3085         & 0.5554          & 0.3798         & 0.9826           & 1.1990            \\
        & $P_g$ [\si{\pu}]        & 0.003706       & 0.06088         & 0.04615        & 0.9868           & 1.9880            \\
        & $Q_g$ [\si{\pu}]        & 0.008996       & 0.09485         & 0.06898        & 0.9687           & 1.9090            \\
        \midrule
        \multirow{4}{*}{\textbf{case9}}
        & $V_m$ [\si{\pu}]        & 6.174e-06      & 0.002485        & 0.001728       & 0.9956           & 0.5519            \\
        & $\delta$ [\si{\degree}] & 0.6066         & 0.7789          & 0.5323         & 0.9943           & 0.5708            \\
        & $P_g$ [\si{\pu}]        & 0.004112       & 0.06413         & 0.05178        & 0.9925           & 1.3800            \\
        & $Q_g$ [\si{\pu}]        & 0.004366       & 0.06608         & 0.05067        & 0.9364           & 2.8470            \\
        \midrule
        \multirow{4}{*}{\textbf{case14}}
        & $V_m$ [\si{\pu}]        & 1.181e-05      & 0.003437        & 0.001882       & 0.9794           & 0.7643            \\
        & $\delta$ [\si{\degree}] & 0.6964         & 0.8345          & 0.5249         & 0.9878           & 0.8566            \\
        & $P_g$ [\si{\pu}]        & 0.007330       & 0.08562         & 0.06335        & 0.9816           & 2.2490            \\
        & $Q_g$ [\si{\pu}]        & 0.01058        & 0.1029          & 0.07391        & 0.9604           & 2.1170            \\
        \midrule
        \multirow{4}{*}{\textbf{case30}}
        & $V_m$ [\si{\pu}]        & 4.063e-06      & 0.002016        & 0.001434       & 0.9901           & 0.5899            \\
        & $\delta$ [\si{\degree}] & 0.4122         & 0.6420          & 0.4702         & 0.9665           & 1.6820            \\
        & $P_g$ [\si{\pu}]        & 0.003126       & 0.05591         & 0.04347        & 0.9848           & 2.0790            \\
        & $Q_g$ [\si{\pu}]        & 0.004686       & 0.06845         & 0.05012        & 0.9894           & 1.1460            \\
        \midrule
        \multirow{4}{*}{\textbf{case\_ieee30}}
        & $V_m$ [\si{\pu}]        & 1.701e-05      & 0.004125        & 0.002349       & 0.9782           & 0.5411            \\
        & $\delta$ [\si{\degree}] & 1.233          & 1.110           & 0.5259         & 0.9781           & 0.7243            \\
        & $P_g$ [\si{\pu}]        & 0.01333        & 0.1154          & 0.05312        & 0.9728           & 1.5310            \\
        & $Q_g$ [\si{\pu}]        & 0.01116        & 0.1056          & 0.07014        & 0.9657           & 1.8230            \\
        \midrule
        \multirow{4}{*}{\textbf{case39}}
        & $V_m$ [\si{\pu}]        & 4.844e-05      & 0.006960        & 0.003952       & 0.9666           & 1.1300            \\
        & $\delta$ [\si{\degree}] & 7.098          & 2.664           & 1.660          & 0.9924           & 0.7471            \\
        & $P_g$ [\si{\pu}]        & 0.04840        & 0.2200          & 0.1571         & 0.9996           & 0.3408            \\
        & $Q_g$ [\si{\pu}]        & 0.1314         & 0.3625          & 0.2275         & 0.9729           & 0.9847            \\
        \midrule
        \multirow{4}{*}{\textbf{case57}}
        & $V_m$ [\si{\pu}]        & 2.840e-05      & 0.005329        & 0.003240       & 0.9957           & 0.5885            \\
        & $\delta$ [\si{\degree}] & 0.7160         & 0.8462          & 0.5790         & 0.9913           & 0.8674            \\
        & $P_g$ [\si{\pu}]        & 0.02938        & 0.1714          & 0.1481         & 0.9963           & 1.1820            \\
        & $Q_g$ [\si{\pu}]        & 0.02278        & 0.1509          & 0.1159         & 0.9815           & 1.8360            \\
        \midrule
        \multirow{4}{*}{\textbf{case118}}
        & $V_m$ [\si{\pu}]        & 4.980e-06      & 0.002232        & 0.001633       & 0.9872           & 1.1450            \\
        & $\delta$ [\si{\degree}] & 2.490          & 1.578           & 1.134          & 0.9943           & 0.8487            \\
        & $P_g$ [\si{\pu}]        & 0.04800        & 0.2191          & 0.1647         & 0.9995           & 0.3796            \\
        & $Q_g$ [\si{\pu}]        & 0.07307        & 0.2703          & 0.1914         & 0.9791           & 0.7890            \\
        \midrule
        \multirow{4}{*}{\textbf{case\_illinois200}}
        & $V_m$ [\si{\pu}]        & 9.892e-06      & 0.003145        & 0.002209       & 0.9791           & 1.1470            \\
        & $\delta$ [\si{\degree}] & 0.8369         & 0.9148          & 0.6255         & 0.9906           & 0.9506            \\
        & $P_g$ [\si{\pu}]        & 0.01969        & 0.1403          & 0.09786        & 0.9992           & 0.4183            \\
        & $Q_g$ [\si{\pu}]        & 0.007698       & 0.08774         & 0.06142        & 0.9444           & 0.8882            \\
        \midrule
        \multirow{4}{*}{\textbf{case300}}
        & $V_m$ [\si{\pu}]        & 1.061e-04      & 0.01030         & 0.005610       & 0.9580           & 0.9966            \\
        & $\delta$ [\si{\degree}] & 15.29          & 3.910           & 2.785          & 0.9905           & 0.9322            \\
        & $P_g$ [\si{\pu}]        & 0.1464         & 0.3826          & 0.2788         & 0.9968           & 1.0730            \\
        & $Q_g$ [\si{\pu}]        & 0.1718         & 0.4145          & 0.2747         & 0.9765           & 0.6868            \\
        \midrule
        \multirow{4}{*}{\textbf{case1354pegase}}
        & $V_m$ [\si{\pu}]        & 8.194e-05      & 0.009052        & 0.006673       & 0.8724           & 2.2770            \\
        & $\delta$ [\si{\degree}] & 12.85          & 3.585           & 2.737          & 0.9944           & 0.8632            \\
        & $P_g$ [\si{\pu}]        & 1.818          & 1.348           & 1.088          & 0.9999           & 0.2359            \\
        & $Q_g$ [\si{\pu}]        & 6.963          & 2.639           & 1.407          & 0.9943           & 0.1620            \\
        \midrule
        \multirow{4}{*}{\textbf{All Systems (Pooled)}}
        & $V_m$ [\si{\pu}]        & 7.038e-05      & 0.008389        & 0.005569       & 0.9586           & 0.8852            \\
        & $\delta$ [\si{\degree}] & 10.59          & 3.254           & 2.292          & 0.9943           & 0.7108            \\
        & $P_g$ [\si{\pu}]        & 0.1682         & 0.4102          & 0.1828         & 0.9998           & 0.0396            \\
        & $Q_g$ [\si{\pu}]        & 3.916          & 1.979           & 0.8633         & 0.9942           & 0.0994            \\
        \bottomrule
    \end{tabular*}
\end{table*}

\begin{table*}[p]
    \centering
    \caption{Unified PowerModelsGAT-AI Performance Across Systems under $N\!-\!0$, $N\!-\!1$, and $N\!-\!2$ Contingencies. Comparison of Mean Squared Error (MSE), Mean Absolute Error (MAE), and Coefficient of Determination ($R^2$) for voltage magnitude, voltage angle, and power generation.}
    \label{tab:unified-n0-n1-n2}
    \renewcommand{\arraystretch}{1.1}
    \setlength{\tabcolsep}{0pt}
    \footnotesize
    \begin{tabular*}{0.95\textwidth}{@{\extracolsep{\fill}}
            l
            l
            S[table-format=1.3e-2, round-mode=figures, round-precision=3, exponent-mode=scientific]
            S[table-format=1.4, round-mode=places, round-precision=4]
            S[table-format=1.4, round-mode=places, round-precision=4]
            S[table-format=1.3e-2, round-mode=figures, round-precision=3, exponent-mode=scientific]
            S[table-format=1.4, round-mode=places, round-precision=4]
            S[table-format=1.4, round-mode=places, round-precision=4]
            S[table-format=1.3e-2, round-mode=figures, round-precision=3, exponent-mode=scientific]
            S[table-format=1.4, round-mode=places, round-precision=4]
            S[table-format=1.4, round-mode=places, round-precision=4]
            @{}}
        \toprule
        &                         & \multicolumn{3}{c}{\textbf{N-0 (Normal)}} & \multicolumn{3}{c}{\textbf{N-1 Contingency}} & \multicolumn{3}{c}{\textbf{N-2 Contingency}} \\
        \cmidrule(lr){3-5} \cmidrule(lr){6-8} \cmidrule(l){9-11}
        \textbf{System} & \textbf{Target}         & {\textbf{MSE}}                               & {\textbf{MAE}}                               & {\textbf{R\textsuperscript{2}}}                             & {\textbf{MSE}} & {\textbf{MAE}} & {\textbf{R\textsuperscript{2}}} & {\textbf{MSE}} & {\textbf{MAE}} & {\textbf{R\textsuperscript{2}}} \\
        \midrule
        \multirow{4}{*}{\textbf{case4gs}}
        & $V_m$ [\si{\pu}]        & 8.439e-07                                    & 0.000726                                     & 0.9972                                       & 3.296e-06      & 0.001300       & 0.9975           & {--}           & {--}           & {--}             \\
        & $\delta$ [\si{\degree}] & 0.05531                                      & 0.1771                                       & 0.9873                                       & 0.2326         & 0.3360         & 0.9944           & {--}           & {--}           & {--}             \\
        & $P_g$ [\si{\pu}]        & 0.006174                                     & 0.05699                                      & 0.9955                                       & 0.007497       & 0.06261        & 0.9951           & {--}           & {--}           & {--}             \\
        & $Q_g$ [\si{\pu}]        & 0.007493                                     & 0.06966                                      & 0.9927                                       & 0.01166        & 0.08261        & 0.9902           & {--}           & {--}           & {--}             \\
        \midrule
        \multirow{4}{*}{\textbf{case5}}
        & $V_m$ [\si{\pu}]        & 1.123e-06                                    & 0.000845                                     & 0.9964                                       & 2.462e-06      & 0.001220       & 0.9940           & 3.641e-06      & 0.001473       & 0.9930           \\
        & $\delta$ [\si{\degree}] & 0.1165                                       & 0.2547                                       & 0.9823                                       & 0.3963         & 0.4570         & 0.9798           & 0.9507         & 0.6975         & 0.9838           \\
        & $P_g$ [\si{\pu}]        & 0.04015                                      & 0.1095                                       & 0.9929                                       & 0.03258        & 0.1112         & 0.9917           & 0.04450        & 0.1570         & 0.9891           \\
        & $Q_g$ [\si{\pu}]        & 0.1462                                       & 0.2553                                       & 0.9933                                       & 0.1544         & 0.3050         & 0.9898           & 0.1324         & 0.2783         & 0.9873           \\
        \midrule
        \multirow{4}{*}{\textbf{case6ww}}
        & $V_m$ [\si{\pu}]        & 1.773e-06                                    & 0.001035                                     & 0.9975                                       & 4.544e-06      & 0.001634       & 0.9968           & 1.085e-05      & 0.002488       & 0.9952           \\
        & $\delta$ [\si{\degree}] & 0.09565                                      & 0.2436                                       & 0.9900                                       & 0.1999         & 0.3328         & 0.9883           & 0.7140         & 0.6143         & 0.9733           \\
        & $P_g$ [\si{\pu}]        & 0.001925                                     & 0.03408                                      & 0.9943                                       & 0.003568       & 0.04566        & 0.9886           & 0.006307       & 0.06320        & 0.9620           \\
        & $Q_g$ [\si{\pu}]        & 0.003978                                     & 0.04681                                      & 0.9828                                       & 0.007937       & 0.06757        & 0.9746           & 0.01709        & 0.1012         & 0.9492           \\
        \midrule
        \multirow{4}{*}{\textbf{case9}}
        & $V_m$ [\si{\pu}]        & 2.590e-06                                    & 0.001256                                     & 0.9967                                       & 1.103e-05      & 0.002369       & 0.9945           & {--}           & {--}           & {--}             \\
        & $\delta$ [\si{\degree}] & 0.2642                                       & 0.3435                                       & 0.9943                                       & 1.070          & 0.7866         & 0.9943           & {--}           & {--}           & {--}             \\
        & $P_g$ [\si{\pu}]        & 0.003355                                     & 0.04708                                      & 0.9938                                       & 0.005018       & 0.05714        & 0.9910           & {--}           & {--}           & {--}             \\
        & $Q_g$ [\si{\pu}]        & 0.002815                                     & 0.04196                                      & 0.9288                                       & 0.006446       & 0.06232        & 0.9245           & {--}           & {--}           & {--}             \\
        \midrule
        \multirow{4}{*}{\textbf{case14}}
        & $V_m$ [\si{\pu}]        & 1.966e-06                                    & 0.001090                                     & 0.9944                                       & 5.298e-06      & 0.001695       & 0.9901           & 3.043e-05      & 0.002986       & 0.9637           \\
        & $\delta$ [\si{\degree}] & 0.1501                                       & 0.3026                                       & 0.9933                                       & 0.4976         & 0.4913         & 0.9908           & 1.540          & 0.8111         & 0.9825           \\
        & $P_g$ [\si{\pu}]        & 0.002690                                     & 0.04052                                      & 0.9935                                       & 0.005799       & 0.05830        & 0.9866           & 0.01433        & 0.09493        & 0.9561           \\
        & $Q_g$ [\si{\pu}]        & 0.006667                                     & 0.06184                                      & 0.9768                                       & 0.007808       & 0.06707        & 0.9669           & 0.01817        & 0.09532        & 0.9354           \\
        \midrule
        \multirow{4}{*}{\textbf{case30}}
        & $V_m$ [\si{\pu}]        & 2.014e-06                                    & 0.001100                                     & 0.9941                                       & 3.901e-06      & 0.001404       & 0.9899           & 6.273e-06      & 0.001795       & 0.9872           \\
        & $\delta$ [\si{\degree}] & 0.2146                                       & 0.3685                                       & 0.9816                                       & 0.4001         & 0.4755         & 0.9668           & 0.6226         & 0.5650         & 0.9528           \\
        & $P_g$ [\si{\pu}]        & 0.002001                                     & 0.03609                                      & 0.9913                                       & 0.003596       & 0.04585        & 0.9827           & 0.003939       & 0.05054        & 0.9774           \\
        & $Q_g$ [\si{\pu}]        & 0.004493                                     & 0.04815                                      & 0.9918                                       & 0.004325       & 0.04825        & 0.9897           & 0.005291       & 0.05422        & 0.9857           \\
        \midrule
        \multirow{4}{*}{\textbf{case\_ieee30}}
        & $V_m$ [\si{\pu}]        & 3.447e-06                                    & 0.001461                                     & 0.9927                                       & 2.036e-05      & 0.002611       & 0.9716           & 3.503e-05      & 0.003472       & 0.9732           \\
        & $\delta$ [\si{\degree}] & 0.1555                                       & 0.3125                                       & 0.9925                                       & 2.648          & 0.6511         & 0.9707           & 1.411          & 0.7250         & 0.9795           \\
        & $P_g$ [\si{\pu}]        & 0.001608                                     & 0.03064                                      & 0.9964                                       & 0.03721        & 0.07899        & 0.9400           & 0.006460       & 0.06310        & 0.9829           \\
        & $Q_g$ [\si{\pu}]        & 0.005966                                     & 0.05906                                      & 0.9818                                       & 0.01827        & 0.08177        & 0.9473           & 0.01172        & 0.07543        & 0.9604           \\
        \midrule
        \multirow{4}{*}{\textbf{case39}}
        & $V_m$ [\si{\pu}]        & 2.408e-05                                    & 0.002983                                     & 0.9809                                       & 3.573e-05      & 0.003595       & 0.9700           & 9.783e-05      & 0.005747       & 0.9502           \\
        & $\delta$ [\si{\degree}] & 4.250                                        & 1.344                                        & 0.9949                                       & 6.528          & 1.579          & 0.9924           & 11.82          & 2.208          & 0.9892           \\
        & $P_g$ [\si{\pu}]        & 0.03420                                      & 0.1301                                       & 0.9997                                       & 0.03827        & 0.1466         & 0.9997           & 0.07883        & 0.2030         & 0.9994           \\
        & $Q_g$ [\si{\pu}]        & 0.07499                                      & 0.1850                                       & 0.9858                                       & 0.1353         & 0.2255         & 0.9661           & 0.2079         & 0.2901         & 0.9602           \\
        \midrule
        \multirow{4}{*}{\textbf{case57}}
        & $V_m$ [\si{\pu}]        & 1.058e-05                                    & 0.002399                                     & 0.9982                                       & 4.346e-05      & 0.003707       & 0.9939           & 4.065e-05      & 0.004159       & 0.9944           \\
        & $\delta$ [\si{\degree}] & 0.3431                                       & 0.4542                                       & 0.9953                                       & 0.7814         & 0.6362         & 0.9907           & 1.336          & 0.7445         & 0.9863           \\
        & $P_g$ [\si{\pu}]        & 0.02739                                      & 0.1514                                       & 0.9964                                       & 0.02920        & 0.1491         & 0.9964           & 0.03982        & 0.1614         & 0.9951           \\
        & $Q_g$ [\si{\pu}]        & 0.01840                                      & 0.1038                                       & 0.9852                                       & 0.02422        & 0.1200         & 0.9797           & 0.02904        & 0.1329         & 0.9772           \\
        \midrule
        \multirow{4}{*}{\textbf{case118}}
        & $V_m$ [\si{\pu}]        & 3.825e-06                                    & 0.001505                                     & 0.9899                                       & 5.186e-06      & 0.001656       & 0.9870           & 6.001e-06      & 0.001745       & 0.9846           \\
        & $\delta$ [\si{\degree}] & 1.427                                        & 0.9167                                       & 0.9964                                       & 2.889          & 1.213          & 0.9935           & 3.194          & 1.277          & 0.9928           \\
        & $P_g$ [\si{\pu}]        & 0.02436                                      & 0.1213                                       & 0.9997                                       & 0.06079        & 0.1857         & 0.9994           & 0.06202        & 0.1950         & 0.9994           \\
        & $Q_g$ [\si{\pu}]        & 0.06688                                      & 0.1858                                       & 0.9806                                       & 0.07327        & 0.1921         & 0.9802           & 0.07919        & 0.1963         & 0.9761           \\
        \midrule
        \multirow{4}{*}{\textbf{case\_illinois200}}
        & $V_m$ [\si{\pu}]        & 6.539e-06                                    & 0.001979                                     & 0.9878                                       & 1.333e-05      & 0.002368       & 0.9673           & 1.193e-05      & 0.002490       & 0.9726           \\
        & $\delta$ [\si{\degree}] & 0.4442                                       & 0.5202                                       & 0.9956                                       & 0.8068         & 0.6556         & 0.9897           & 1.927          & 0.8651         & 0.9763           \\
        & $P_g$ [\si{\pu}]        & 0.02026                                      & 0.09325                                      & 0.9993                                       & 0.01691        & 0.1024         & 0.9992           & 0.02486        & 0.1093         & 0.9988           \\
        & $Q_g$ [\si{\pu}]        & 0.008028                                     & 0.06200                                      & 0.9498                                       & 0.007587       & 0.06124        & 0.9394           & 0.006899       & 0.05985        & 0.9357           \\
        \midrule
        \multirow{4}{*}{\textbf{case300}}
        & $V_m$ [\si{\pu}]        & 1.027e-04                                    & 0.005389                                     & 0.9563                                       & 1.106e-04      & 0.005837       & 0.9607           & 1.060e-04      & 0.005672       & 0.9559           \\
        & $\delta$ [\si{\degree}] & 10.79                                        & 2.487                                        & 0.9924                                       & 21.69          & 3.304          & 0.9883           & 14.07          & 2.587          & 0.9905           \\
        & $P_g$ [\si{\pu}]        & 0.1139                                       & 0.2527                                       & 0.9972                                       & 0.2134         & 0.3293         & 0.9957           & 0.1061         & 0.2521         & 0.9977           \\
        & $Q_g$ [\si{\pu}]        & 0.1603                                       & 0.2686                                       & 0.9783                                       & 0.1892         & 0.2806         & 0.9746           & 0.1676         & 0.2768         & 0.9760           \\
        \midrule
        \multirow{4}{*}{\textbf{case1354pegase}}
        & $V_m$ [\si{\pu}]        & 8.222e-05                                    & 0.006702                                     & 0.8704                                       & 7.873e-05      & 0.006557       & 0.8789           & 8.487e-05      & 0.006746       & 0.8687           \\
        & $\delta$ [\si{\degree}] & 12.71                                        & 2.704                                        & 0.9947                                       & 13.44          & 2.847          & 0.9944           & 12.42          & 2.675          & 0.9935           \\
        & $P_g$ [\si{\pu}]        & 1.711                                        & 1.058                                        & 0.9999                                       & 2.522          & 1.302          & 0.9998           & 1.239          & 0.9063         & 0.9999           \\
        & $Q_g$ [\si{\pu}]        & 6.791                                        & 1.406                                        & 0.9943                                       & 7.137          & 1.414          & 0.9941           & 6.987          & 1.397          & 0.9943           \\
        \bottomrule
    \end{tabular*}
\end{table*}

\FloatBarrier
\clearpage
\section{Detailed Continual Learning Results}
\label{app:continual}

This appendix reports the per-system knowledge-loss values after fine-tuning (Table~\ref{tab:continual-detail} and Table~\ref{tab:continual-physical}). Bar-chart visualizations of the same results are provided in the Supplementary Material (Figs.~S15 and S16).

\begin{table*}[!p]
    \centering
    \caption{Per-System Knowledge Loss ($\mathcal{K}_{\tau}^{\text{NMAE}}$) After Fine-Tuning on \code{case1354pegase}. Per-system values are reported for Naive and EWC+Replay across all base systems.}
    \label{tab:continual-detail}
    \renewcommand{\arraystretch}{1.15}
    \setlength{\tabcolsep}{0pt}
    \footnotesize
    \begin{tabular*}{0.95\textwidth}{@{\extracolsep{\fill}} l S[table-format=2.2] S[table-format=1.2] S[table-format=3.2] S[table-format=1.2] S[table-format=4.2] S[table-format=1.2] S[table-format=3.2] S[table-format=1.2] @{}}
        \toprule
        \textbf{System}                                                   &
        \multicolumn{2}{c}{$\mathcal{K}_{V_m}^{\text{NMAE}}$}    &
        \multicolumn{2}{c}{$\mathcal{K}_{\delta}^{\text{NMAE}}$} &
        \multicolumn{2}{c}{$\mathcal{K}_{P_g}^{\text{NMAE}}$}    &
        \multicolumn{2}{c}{$\mathcal{K}_{Q_g}^{\text{NMAE}}$}                                                                              \\
        &
        {\textbf{Naive}}                                                  & {\textbf{EWC+Replay}} &
        {\textbf{Naive}}                                                  & {\textbf{EWC+Replay}} &
        {\textbf{Naive}}                                                  & {\textbf{EWC+Replay}} &
        {\textbf{Naive}}                                                  & {\textbf{EWC+Replay}}                                                          \\
        \midrule
        \textbf{case4gs}                                                  & 19.53          & 0.05 & 258.30 & -0.05 & 1433.74 & 1.14 & 40.41  & 1.25 \\
        \textbf{case5}                                                    & 9.91           & 0.69 & 219.00 & 0.04  & 721.58  & 1.21 & 12.90  & 0.80 \\
        \textbf{case6ww}                                                  & 18.37          & 0.24 & 245.35 & 0.20  & 2887.41 & 1.62 & 121.43 & 1.88 \\
        \textbf{case9}                                                    & 10.38          & 0.21 & 41.72  & 0.10  & 1493.63 & 1.58 & 110.11 & 3.20 \\
        \textbf{case14}                                                   & 7.50           & 0.29 & 59.09  & 0.14  & 2912.61 & 2.26 & 303.94 & 2.32 \\
        \textbf{case30}                                                   & 12.57          & 0.22 & 406.13 & -0.03 & 7737.94 & 1.61 & 47.59  & 1.64 \\
        \textbf{case\_ieee30}                                             & 6.43           & 0.20 & 99.53  & 0.15  & 3280.43 & 2.23 & 287.24 & 2.27 \\
        \textbf{case39}                                                   & 8.73           & 0.31 & 33.42  & 0.12  & 322.10  & 0.15 & 12.68  & 0.67 \\
        \textbf{case57}                                                   & 22.14          & 0.24 & 111.06 & 0.24  & 824.90  & 1.79 & 239.30 & 2.22 \\
        \textbf{case118}                                                  & 9.29           & 0.29 & 42.34  & 0.12  & 208.89  & 0.37 & 23.71  & 0.79 \\
        \textbf{case\_illinois200}                                        & 13.62          & 0.25 & 116.10 & 0.08  & 292.57  & 0.29 & 31.73  & 0.91 \\
        \textbf{case300}                                                  & 5.58           & 0.14 & 26.72  & 0.32  & 586.10  & 0.76 & 14.46  & 0.57 \\
        \bottomrule
    \end{tabular*}
\end{table*}

\begin{table*}[!p]
    \centering
    \caption{Per-System Knowledge Loss ($\mathcal{K}_{\tau}^{\text{MAE}}$) in Physical Units After Fine-Tuning on \code{case1354pegase}. Per-system values are reported for Naive and EWC+Replay across all base systems.}
    \label{tab:continual-physical}
    \renewcommand{\arraystretch}{1.15}
    \setlength{\tabcolsep}{0pt}
    \footnotesize
    \begin{tabular*}{0.95\textwidth}{@{\extracolsep{\fill}} l S[table-format=2.2] S[table-format=1.2] S[table-format=3.2] S[table-format=1.2] S[table-format=3.2] S[table-format=1.2] S[table-format=2.2] S[table-format=1.2] @{}}
        \toprule
        \textbf{System}                                                               &
        \multicolumn{2}{c}{$\mathcal{K}_{V_m}^{\text{MAE}}$ [\si{\pu}]}      &
        \multicolumn{2}{c}{$\mathcal{K}_{\delta}^{\text{MAE}}$ [\si{\degree}]} &
        \multicolumn{2}{c}{$\mathcal{K}_{P_g}^{\text{MAE}}$ [\si{\pu}]}      &
        \multicolumn{2}{c}{$\mathcal{K}_{Q_g}^{\text{MAE}}$ [\si{\pu}]}                                                                              \\
        &
        {\textbf{Naive}}                                                              & {\textbf{EWC+Replay}} &
        {\textbf{Naive}}                                                              & {\textbf{EWC+Replay}} &
        {\textbf{Naive}}                                                              & {\textbf{EWC+Replay}} &
        {\textbf{Naive}}                                                              & {\textbf{EWC+Replay}}                                                        \\
        \midrule
        \textbf{case4gs}                                                              & 0.04           & 0.00 & 97.77  & -0.02 & 83.87  & 0.07 & 2.43  & 0.08 \\
        \textbf{case5}                                                                & 0.01           & 0.00 & 92.95  & 0.02  & 72.30  & 0.12 & 3.65  & 0.23 \\
        \textbf{case6ww}                                                              & 0.05           & 0.00 & 77.70  & 0.06  & 67.03  & 0.04 & 4.39  & 0.07 \\
        \textbf{case9}                                                                & 0.03           & 0.00 & 38.91  & 0.10  & 56.05  & 0.06 & 1.96  & 0.06 \\
        \textbf{case14}                                                               & 0.02           & 0.00 & 36.21  & 0.08  & 82.02  & 0.06 & 10.61 & 0.08 \\
        \textbf{case30}                                                               & 0.03           & 0.00 & 113.55 & -0.01 & 161.79 & 0.03 & 2.08  & 0.07 \\
        \textbf{case\_ieee30}                                                         & 0.03           & 0.00 & 72.26  & 0.11  & 113.83 & 0.08 & 11.05 & 0.09 \\
        \textbf{case39}                                                               & 0.03           & 0.00 & 74.27  & 0.27  & 148.48 & 0.07 & 2.93  & 0.15 \\
        \textbf{case57}                                                               & 0.12           & 0.00 & 74.13  & 0.16  & 103.31 & 0.22 & 15.10 & 0.14 \\
        \textbf{case118}                                                              & 0.01           & 0.00 & 56.57  & 0.17  & 90.67  & 0.16 & 5.75  & 0.19 \\
        \textbf{case\_illinois200}                                                    & 0.03           & 0.00 & 76.40  & 0.05  & 68.45  & 0.07 & 2.19  & 0.06 \\
        \textbf{case300}                                                              & 0.03           & 0.00 & 79.82  & 0.95  & 152.26 & 0.20 & 5.78  & 0.23 \\
        \bottomrule
    \end{tabular*}
\end{table*}

\FloatBarrier
\clearpage
\section*{Conflict of Interest}
None of the authors have a conflict of interest to disclose.

\bibliographystyle{IEEEtran}
\bibliography{refs}

\begin{IEEEbiography}[{\includegraphics[width=1in,height=1in,clip,keepaspectratio]{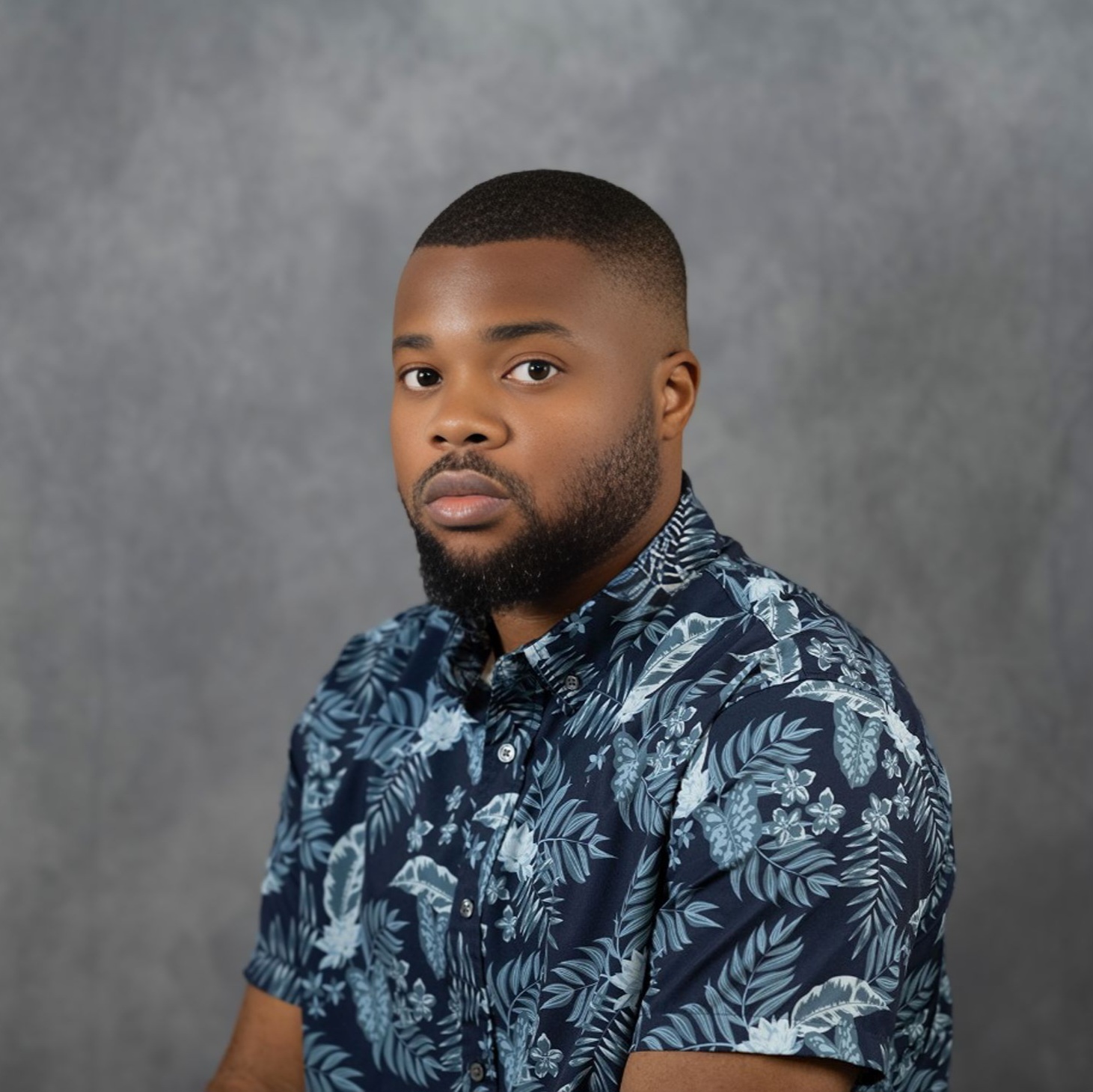}}]{Chidozie Ezeakunne}
received the M.Sc. degree in Physics from the University of Central Florida in 2025, where he is currently pursuing the Ph.D. degree in Physics. He is a Graduate Intern with Los Alamos National Laboratory. His research interests include scientific machine learning, graph-based modeling, computational physics, and data-driven methods for complex physical and engineering systems.
\end{IEEEbiography}

\begin{IEEEbiography}[{\includegraphics[width=1in,height=1in,clip,keepaspectratio]{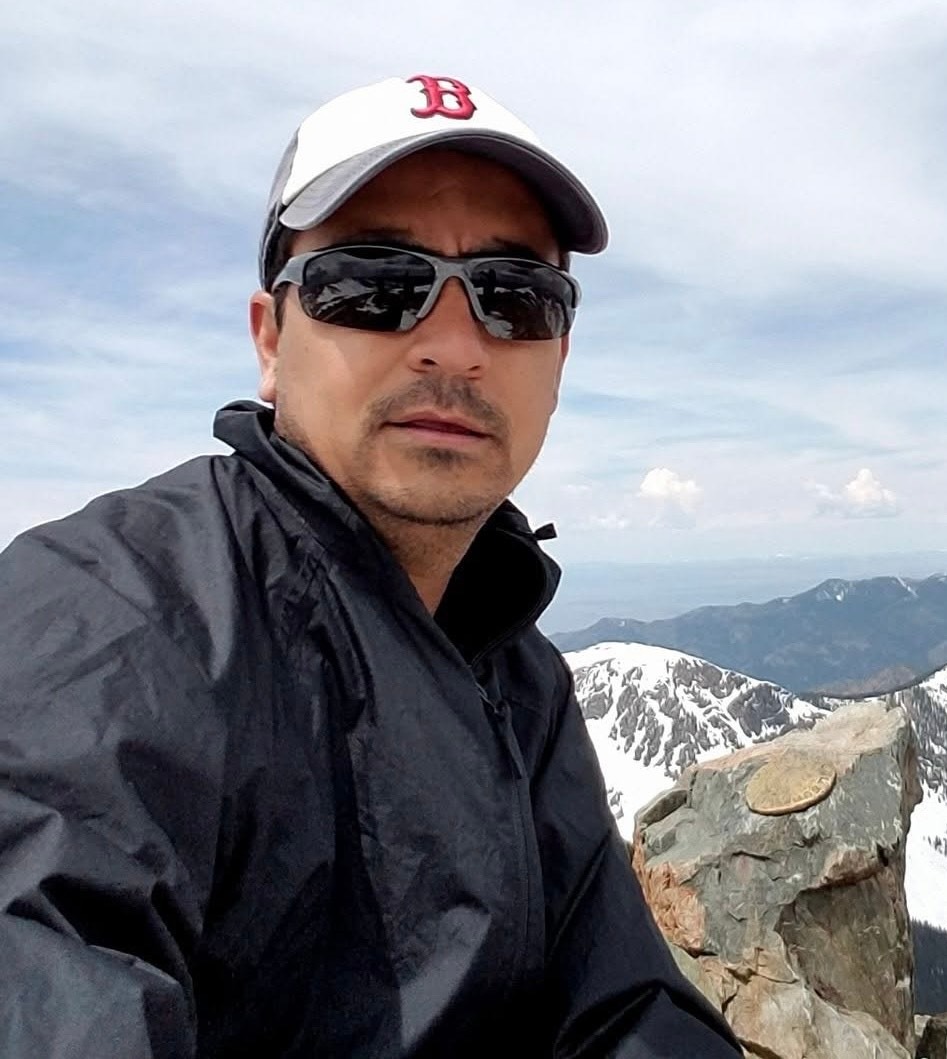}}]{JOSE E. TABAREZ}
received his Ph.D. from New Mexico State University in Electrical Engineering in 2019. At New Mexico State University he was an EMUP and Sandia National Laboratories fellow. He is currently a research and development engineer at Los Alamos National Laboratory. His focus is on power system optimization and analysis, and modeling effects on power systems from geomagnetic disturbances.
\end{IEEEbiography}

\begin{IEEEbiography}[{\includegraphics[width=1in,height=1in,clip,keepaspectratio]{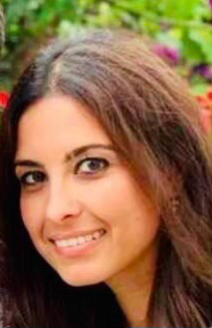}}]{Reeju Pokharel}
is a staff scientist in the Materials Science and Technology Division at Los Alamos National Laboratory. She received her Ph.D. in materials science from Carnegie Mellon University in 2013. Her research focuses on data science for 3D imaging and integrating physics constraints into generative machine learning models to provide real-time feedback during dynamic imaging experiments at light sources.
\end{IEEEbiography}

\begin{IEEEbiography}[{\includegraphics[width=1in,height=1in,clip,keepaspectratio]{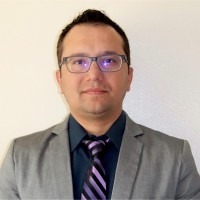}}]{ANUP PANDEY}
received his M.S. and Ph.D. degrees in Physics from Ohio University, Athens, OH, in 2014 and 2017, respectively. He served as a postdoctoral research associate at Oak Ridge National Laboratory from 2017 to 2019. His research interests focus on the application of artificial intelligence and physics-informed machine learning to critical infrastructure systems.
\end{IEEEbiography}

\EOD

\end{document}